%% file: ms.tex
\documentclass[12pt,preprint]{aastex}

\begin{document}

\title{Catalog of fundamental mode RR Lyrae stars in the Galactic bulge from the Optical Gravitational Lensing Experiment}
\author{Matthew J. Collinge\altaffilmark{1}, Takahiro Sumi\altaffilmark{1}, Daniel Fabrycky\altaffilmark{1}}
\altaffiltext{1}{Princeton University Observatory, Princeton, New Jersey 08544, USA}


\begin{abstract}
\end{abstract}

We present a catalog of 1888 distinct fundamental-mode RR Lyrae stars detected 
in the Galactic bulge fields of the second phase of the Optical Gravitational 
Lensing Experiment (OGLE), covering an area near 11~deg$^2$. 
These stars were selected 
primarily according to their light curve morphologies. The catalog includes 
basic parameters of the light curves, significant frequencies detected 
close to the main pulsation frequencies (characteristic of the Blazhko effect), 
$V-I$ colors at minimum light (for most stars), and information useful 
for assessing the quality of the data for each star.
We detect a high rate of incidence of the Blazhko phenomenon (at least 27.6\%), 
including an unprecedentedly large proportion of stars with symmetrical 
frequency triplets, which we attribute to the new and sensitive method we 
employ to search for them.
We find that the minimum light $V-I$ color (useful as a reddening indicator) 
grows slowly redder with increasing period and exhibits an unexpectedly large 
star-to-star scatter of approximately 0.07~mag.
We use this color to evaluate the zero-point accuracy of 
the reddening map of the Galactic bulge derived from OGLE data, 
and find that in addition to probable low-level random errors or resolution effects 
(responsible for much of the scatter), the map may systematically 
over-represent $E(V-I)$ by approximately 0.05~mag in most fields.
While the conclusion about the reddening zero points is somewhat tentative, 
we have reasonably robust evidence that the RR Lyrae-to-red clump color 
separation is larger by 0.05--0.08~mag in the bulge than locally; at 
a minimum this sounds a cautionary note about the use of these stars 
for reddening determinations until the effect is better constrained.
We consider the RR Lyrae constraint on the Galactocentric distance, but 
our uncertainty about the absolute magnitude calibration 
and possible errors in the extinction determinations leave significant 
flexibility in the result. However, in contrast to previous results, we 
robustly detect the signature of the Galactic bar in the RR Lyrae population 
within the inner $\pm 3^{\circ}$ of longitude, and we highlight the 
apparent differences between the structures traced by the red clump giants 
and the more metal-poor RR Lyrae stars.



\section{Introduction}

The photometric data sets obtained in the course of large microlensing surveys 
provide an abundance of opportunities to carry out scientific investigations not 
related to lensing. In one such project, Sumi~(2004; S04 hereafter) presented 
extinction maps of the Galactic bulge fields observed in the Optical Gravitational 
Lensing Experiment (OGLE) during its second phase. These fields range in Galactic 
longitude between approximately $-11$~deg$<l<11$~deg and cover an area close 
to 11 deg$^2$. The purpose of this effort was to enable studies of Galactic 
structure with the same data set. 

The method employed by S04 was to determine the location of the 
red clump (RC) giants in the $V-I, I$ color-magnitude diagram for a given 
sub-region of sky (presumably with homogeneous or nearly-homogeneous extinction), 
and use the observed color of the RC as a measure of the reddening. By locating 
the RC in the color-magnitude diagrams for sub-regions with different amounts 
of extinction, assuming the RC to be constant in luminosity as well as color, 
the reddening slope can be measured and extrapolated to 
zero reddening to obtain the total extinction (with some calibration).
As suggested by \citet{popo00} and confirmed by \citet{udal03}, 
S04 also found that the measured reddening slopes $1.9 \la R_{VI} \la 2.1$ were 
significantly flatter than the ``standard'' value of about 2.5. Here 
$R_{VI}\equiv A_V/E(V-I)$ is the usual ratio of total to selective absorption. 
This ``anomalously'' shallow extinction slope provides an explanation for previous 
claims (e.g., \citealt{stut99}) that extinction corrected colors (derived 
according to a standard reddening law) of stars in the bulge were redder 
than their local counterparts.

The minimum light colors of fundamental mode RR Lyrae (RRab)
stars have long been recognized as 
useful reddening measures \citep{stur66} because of their low intrinsic 
dispersions and weak dependences on metallicity.
The $V-I$ color was suggested to be a particularly good reddening standard 
by \citet{mate95} based on the data of \citet{liu89}. Recently,  
\citet{day02} and \citet{guld05} have studied the colors of field 
RR Lyrae stars and found the intrinsic minimum light color to be 
$(V-I)_{0, ml} = 0.58\pm 0.02$ (consistent with the value used by 
\citealt{mate95}), with a scatter of 0.024~mag.

S04 considered the possibility that the intrinsic colors of RC stars 
might vary somewhat from field to field, due to a possible weak dependence on 
metallicity; in order to test this he compiled a list of RRab stars. 
In S04's original analysis, the precision 
of the $V-I$ colors of the RR Lyrae stars was limited, because the available 
$V$-band measurements were averages of small numbers of observations taken 
at random (and unknown) phases. In this work we present and analyze the 
catalog of RRab stars in the Galactic bulge.
Using individual $V$-band time series, we extract measurements of $(V-I)_{ml}$ 
for bulge RR Lyrae stars and use them to evaluate the 
field-to-field zero point accuracy of S04's extinction map, 
among other applications.

In \S\ref{sec:data} we discuss the OGLE data, in \S\ref{sec:selection} we present 
the sample selection procedure, and in \S\ref{sec:analysis} we analyze the light 
curves. In \S\ref{sec:vmi} we extract and analyze the $(V-I)_{ml}$ colors. 
In \S\ref{sec:disc} 
we discuss the extinction zero points, distance scale, and geometry of the 
inner Galaxy, and in \S\ref{sec:conc} we summarize our 
main conclusions.


\section{OGLE observations and data sets}
\label{sec:data}

The data used in the present work were collected during the second 
phase of the OGLE project (OGLE-II hereafter). Details of the instrumental 
setup, calibration, and data pipeline can be found in Udalski et al.\ (2002; U02 
hereafter) and references therein, who presented $BVI$ photometric maps of the 
OGLE-II Galactic bulge fields. In addition to the photometric maps, we primarily 
make use of the $I$-band time series obtained through difference image 
analysis (DIA; \citealt{alar98,alar00}) as presented by \citet{szym05}, though 
our analysis is based on a somewhat older version of the reductions. 
DIA photometry for BUL\_SC28 were not available in these reductions, so our 
analysis is confined to the remaining 48 OGLE-II bulge fields.
We also use the $V$-band time series, reduced via the standard 
(non-DIA) OGLE pipeline and kindly provided by the OGLE team. 
We transform the OGLE $V$ and $I$ magnitudes 
to the standard Johnson-Kron-Cousins system as prescribed by U02, after 
correcting for the low-level systematic flat-fielding errors mentioned 
in \S5 of that work (M. Szyma{\'n}ski 2004, private communication).

Two issues that can affect the utility of a photometric data set are the 
quality of the photometric error estimates and the choice of ``good'' frames. 
We adopt the following approach to these issues. On each 
frame, we construct a distribution of flux offsets for constant stars 
(chosen to have $\chi^2$ per degree of freedom less than some threshold for 
a constant fit). We reject frames for which this distribution is significantly 
non-Gaussian, again based on a $\chi^2$ threshold. 
For the remaining frames, we determine a scaling factor as 
a function of magnitude by which we multiply the photometric error estimates 
to allow accurate application of $\chi^2$ fitting statistics. These 
procedures apply only to the $I$-band data; the much smaller number of 
$V$-band frames does not permit the same approach. Thus no error-scaling 
is performed for the $V$-band data points, and the choice of good frames is 
based on the standard OGLE data quality flags.

Since our analysis is based on non-public data, we performed several 
consistency checks to ensure that there were no major calibration problems 
as compared with published reductions of the same data sets. 
We compared the mean RR Lyrae magnitudes obtained from the $V$-band time series 
with the same quantity as presented by U02. In the majority of cases, 
the mean $V$ magnitudes we obtained agree with those presented by U02 
to better than 1\%; however, there are also tails to the $\Delta V_{Up}$ 
(the subscript $Up$ indicates a comparison between values from U02 
and those from the present work) distribution 
that extend out to approximately $\pm 0.2$~mag. 
These cases probably correspond to differences 
in the specific observations that were chosen to calculate the mean magnitudes; 
we applied a $5\sigma$ filter before calculating the mean magnitudes for 
consistency with U02, but our implementation may differ slightly 
from that of U02, and we cannot guarantee that the choice of ``good'' 
frames is identical (a 
step that is performed prior to applying the $5\sigma$ filter). Since the 
$V$-band light curves typically contain only 5--15 observations and typical 
amplitudes of RR Lyrae stars are several tenths of a magnitude or more, 
adding or subtracting a data point here and there can make a significant 
difference in the mean magnitude. Therefore we do not regard the width of the 
$\Delta V_{Up}$ distribution as a matter for serious concern, especially since 
there is no evidence of a significant systematic offset between the 
two sets of magnitudes (mean $\Delta V_{Up} \ll 0.01$). For the sake of caution, 
we exclude stars with $|\Delta V_{Up}|>0.1$ from subsequent analysis of the 
$V-I$ colors.

We also compared the mean RR Lyrae magnitudes obtained from the DIA $I$-band 
time series against those presented by U02. Again we found no significant 
systematic offsets, and the root-mean-square (rms) difference in magnitudes 
for objects with $|\Delta I_{Up}|<0.2$ was $\Delta I_{Up,rms} = 0.035$. This is 
approximately the expected level of agreement. 
Although the $\Delta I_{Up}$ distribution is somewhat narrower 
than the $\Delta V_{Up}$ distribution 
(as expected since the number of observations in the $I$-band is much larger, 
and hence the impact of averaging over different subsets of the data is much 
smaller), there is still a slight tail of objects extending out to $\pm 1$~mag. 
This can arise naturally because of crowding. If a star is blended with another 
star at close to the resolution limit of the survey, then it may be possible 
to separate the two on some frames but not others. Thus it may occur that the 
baseline magnitude obtained from DIA photometry may differ from the value 
obtained from standard point-spread-function (PSF) photometry, and the 
effect may have either sign. We suspect that this is the reason for objects with 
large values of $|\Delta I_{Up}|$. In any case such objects must be treated as 
suspicious; in the analysis of the $V-I$ colors we exclude stars with 
$|\Delta I_{Up}|>0.1$.


\section{Selection of fundamental mode RR Lyrae}
\label{sec:selection}

S04 selected 1961 RRab candidates from the OGLE-II catalog of variable 
stars in the Galactic bulge \citep{wozn02} using the method of \citet{alar96}. 
Periods were determined with the phase dispersion minimization technique 
\citep{stel78}, and the $I$-band light curves were decomposed as Fourier 
series of five harmonics. (For clarity, we remark that this 
modeling utilized the $I$-band data as presented by \citealt{wozn02}, which 
differs from the reductions used elsewhere in the present work.) 
Among variables with periods shorter than 0.9~days, RRab 
candidates were selected as a clump in the $\phi_{21}$, $R_{21}$ plane. 
To avoid confusion arising from the window function imposed by the OGLE 
observing cadence, stars with periods in the range $0.4985<P<0.5001$~day 
were excluded.
According to convention, $\phi_{21}\equiv \phi_2-2\phi_1$ and 
$R_{21}\equiv A_2/A_1$, where $\phi_i$ and $A_i$ are the phase (in radians) and 
amplitude of harmonic $i$ in the Fourier series. Figure~10 of S04 
shows the ellipse used to select the RRab candidates; the 
ellipse is centered on $(4.5~rad, 0.43)$ with semi-major axis $a=0.8$ and 
semi-minor axis $b=0.17$, and the angle between the horizontal and the major 
axis is $-10$~deg.

From the list of 1961 RRab candidates, we rejected redundant entries (5 stars), 
stars that had less than 25 data points in the $I$-band (26 stars, including 
all 13 RRab candidates in BUL\_SC28), 
one star that displayed an RR-Lyra-like light curve because it was 
blended with a true RR Lyra star, and 16 more stars that 
showed clear non-RR Lyrae light curves (e.g., constant light curves). 
Of the remaining 1913 RRab candidates, not all are unique; there are 
25 stars that appear in the list twice because they lie in overlap regions 
between two fields. Because the number of 
multiply-identified stars is small and the duplicate data sets consist 
of independent measurements, we treat the duplicate entries independently 
in the analysis that follows. These stars provide a useful cross-check 
of the selection procedure and the overall photometric precision. 

\begin{figure}
\epsscale{1.0}
\plotone{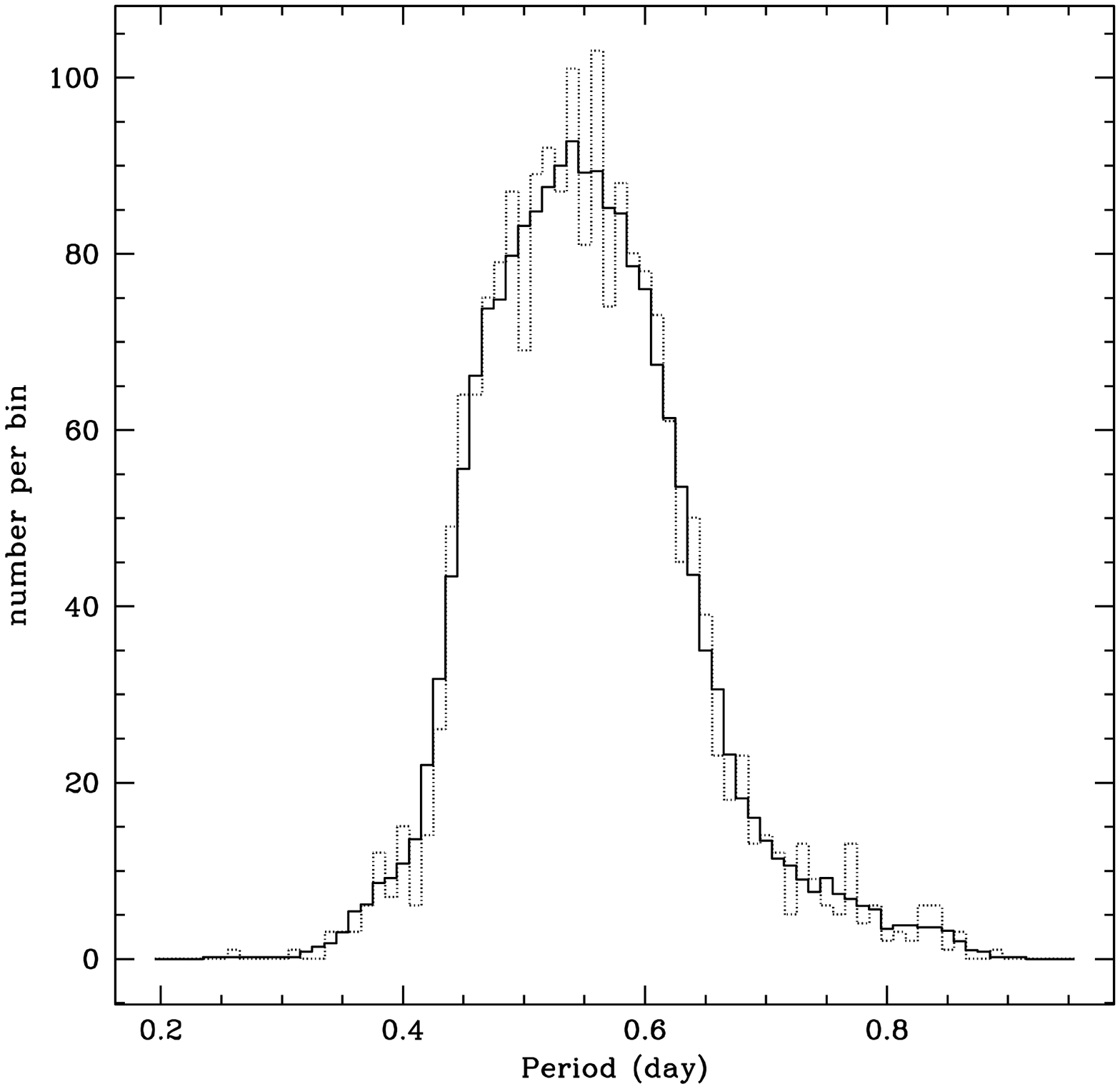}
\figcaption{
Period distribution of the 1913 stars in our sample. The light histogram is the 
observed distribution with a bin size of 0.01~day. The dark histogram is the 
result of smoothing the observed distribution with a five-bin boxcar filter.
\label{fig:phist}}
\end{figure}

Figure~\ref{fig:phist} shows the distribution of periods. One can see a small dip 
near 0.5~day that results from the selection requirement mentioned above. 
The modal value of period indicated by the observed distribution is 0.56~day; 
after smoothing with a 5-bin boxcar filter, the peak shifts to 
0.54~day, which appears somewhat more satisfactory to the eye. The mean period 
is intermediate between the two peaks at 0.552~day. This value indicates a 
similarity between RR Lyrae stars in the bulge and those found in Oosterhoff 
type~I globular clusters (e.g., \citealt{smit95}), as expected if the RR Lyrae 
stars in the bulge are on average relatively metal rich. Our mean period is 
consistent with the 0.554~day reported by \citet{mize03}, who also analyzed 
bulge RR Lyrae stars detected in OGLE-II. It is somewhat 
lower than the mean LMC RRab periods of 0.573~day obtained by \citet{sosz03} 
and 0.583~day obtained by \citet{alco96}, and the mean SMC RRab period 
of 0.589~day obtained by \citet{sosz02}.

We present the list of 1913 RRab candidates (including the 25 duplicates) 
in Table~1, along with catalog 
identification numbers, mean photometry and equatorial coordinates from U02. 
The table also contains flags to warn of unreliable photometry, labels indicating 
any multi-periodic nature as described in \S\ref{sec:freq}, and 
periods, amplitudes, $I$ magnitudes at mean flux, $V-I$ colors at minimum light, 
and $\chi^2$ values derived from the modeling procedure discussed 
in \S\ref{sec:model}. The 43 unique stars rejected from the sample are listed 
in Table~2 for completeness. 


\section{Analysis of the light curves}
\label{sec:analysis}


\subsection{Frequency analysis}
\label{sec:freq}

Many RR Lyrae stars display modulations of their light curves that are 
periodic or nearly so, with typical modulation periods in the range of tens 
to hundreds of days; these are the Blazhko stars (e.g., \citealt{smit95}).
In the frequency domain, the Blazhko effect can manifest itself 
as an additional significant peak or peaks close to the main pulsation frequency, 
with the difference in frequency corresponding to the modulation frequency. 
Two major subclasses are recognized in the literature: those having a single 
additional significant frequency, and those having two additional significant 
frequencies, one on either side of the main frequency.
These two respective classes are sometimes termed BL1 and BL2, and it is often 
required that a frequency triplet be evenly spaced in order to receive the 
BL2 designation (e.g., \citealt{mize03}). There are also Blazhko stars that 
do not conform to either class, having more than two or two unevenly spaced 
additional frequencies.

In order to identify Blazhko stars in our sample, we first pre-whiten the 
light curves by subtracting best-fit Fourier series of eight harmonics of the 
main pulsation period. (For details about Fourier analysis see 
\S\ref{sec:model}.) Next we extract the frequency spectrum by running 
the CLEAN algorithm \citep{robe87} on the 
pre-whitened light curves for 200 iterations, with a gain of 0.5, on the frequency 
band [0--5~day$^{-1}$] with a frequency resolution of $0.125/T$, where $T$ is the 
overall time baseline of the observations. We sort the peaks and take the amplitude 
of the 15th largest peak (${\cal{A}}_{15}$) to be indicative of the noise level. 
The narrower the bandwidth searched, the less likely we are to find a spurious 
peak at a given amplitude. Since the different categories of secondary 
periodicities have different bandwidths, we determined a set of significance 
thresholds via a Monte-Carlo method described below. In what follows, we 
define the ``Blazhko range'' to be frequencies within 0.1~day$^{-1}$ of, but 
separated by greater than $1/T$ from, the main pulsation frequency. We 
exclude frequencies that are plausibly aliases (e.g., near-integer frequencies) 
from the classification scheme but include these frequencies in our noise 
estimates; this filter is required to guard against false positives induced 
by aliasing of low frequency noise, which our Monte-Carlo method overlooks. 
Only a small number of stars are classified more conservatively because of 
this filter.

To qualify as a BL1, a star 
must have a single peak in the CLEANed spectrum with amplitude 
${\cal{A}} > 2.4 {\cal{A}}_{15}$ 
in the Blazhko range. In order to be 
considered a BL2, a star must have two frequencies within the Blazhko range
(one on each side of the main pulsation frequency), with offsets from the 
main frequency that are identical to 
within $3.0/T$, and whose amplitudes both satisfy 
${\cal{A}} > 1.1 {\cal{A}}_{15}$. (The frequency 
condition follows the definition of an equidistant triplet found in 
\citealt{alco03}). A star with more than one peak satisfying 
${\cal{A}} > 1.6 {\cal{A}}_{15}$ 
within the Blazhko range but not satisfying the equidistant 
triplet condition is called ``BL?''. BL2 and BL?\ classifications supersede 
BL1; a star satisfying the BL2 condition with an additional peak that satisfies 
the BL?\ condition is called ``BL2+?''. Finally, a star with a peak 
satisfying ${\cal{A}} > 1.39 {\cal{A}}_{15}$ and within $1/T$ of the main frequency 
is labeled PC for period change, following the discussion of \citet{alco03}.  
Such peaks may be indicative of a Blazhko period that is longer than the time 
baseline of the data set, but sometimes a star meets BL criteria as well, causing 
it to be labeled, for example, BL1+PC. 

To refine the amplitude measurements for the additional significant 
frequencies, we used these frequencies to initialize a 
non-linear solver (the Levenberg-Marquardt algorithm; \citealt{pres89}); 
the resulting frequencies and amplitudes are reported in Table~3. 
The best-fitting model was subtracted from the light curve and the amplitude of 
the highest remaining peak within 0.1~day$^{-1}$ of the main frequency is also 
reported in Table~1. We do not consider this peak significant; it is reported 
to quantify the level of the noise for completeness studies. For BL1 stars we 
also report (in Table~3) the amplitude of the highest peak within the 
equidistant triple limits in frequency, which can be used to study the important 
question of whether the BL2 phenomenon makes a continuous transition to BL1.

The amplitude thresholds cited above were determined by a Monte-Carlo realization 
of the data set for each star, in which the magnitude residuals from the 
pre-whitening step were randomly swapped among the observation times. Next, the 
CLEAN algorithm was run and the amplitude thresholds of the routine to label the 
peaks were varied to produce an acceptable false alarm rate such that about 1\% 
of the stars in each of the categories BL1, BL2, BL?\ and PC 
are likely to be spurious. 

In addition to the expected BL1, BL2 and PC stars and stars displaying 
BL?\ behavior, our procedure allows us to identify several stars 
with double symmetrical frequency triplets (two sets of equidistant sidebands) 
in the Blazhko range. These stars have two pairs 
of peaks that both satisfy the BL2 conditions explained above.
A similar phenomenon has been observed at least once before \citep{lacl04}. 
We identify these 10 stars 
in Table~1 with the label BL2x2. While most of these stars exhibit a 
very closely spaced pair of frequencies on each side of the main pulsation 
frequency, which may indicate a BL2 phenomenon with unstable modulation 
frequency, there are also several interesting cases of stars showing two 
well-separated frequencies on each side of the main pulsation frequency. 
In contrast to expectation (e.g., \citealt{alco03}), none of these 
frequency structures form evenly spaced quintuplets.

The number of unique stars in each of our categories is 
(BL1, BL2, BL?, BL2+?, BL2x2, PC) = (167, 282, 28, 37, 10, 119) = (8.8\%, 
14.9\%, 1.5\%, 2.0\%, 0.5\%, 6.3\%). While we have confidence 
in the robustness of our detections of additional significant frequencies, 
many are close to the limits of our ability to detect them, which indicates 
that we may still be missing a significant proportion of Blazhko stars. 
Among the 25 stars detected twice in overlapping fields, 10 of which are 
classified as having some type of BL/PC behavior, eight of the duplicate 
data sets receive different classifications. Two of these are cases of 
stars switching from one BL category to another because of the detection of 
an additional peak in one of the data sets (these stars contribute 
twice to the numbers listed above); the remaining six stars are simply not 
recognized as BL/PC in one of the data sets. In all cases, the 
amplitudes of the extra peaks are close to the noise levels of the data sets 
in which the extra peaks were not detected. Thus there is no conflict 
with our estimate of completeness for each data set, but this underscores 
the fact that more Blazhko stars remain to be discovered in our sample.

Despite the likely incompleteness of our Blazhko classifications, we have 
detected a very high rate of incidence of the Blazhko phenomenon compared 
to previous studies. Overall 
$522/1888=27.6$\% of our unique stars display some sort of Blazhko behavior 
(not including stars with only PC labels) and an additional 
4.8\% are classified solely as PC stars. The former number is somewhat 
larger than the total Blazhko incidence rates of $\sim 23$\% reported by 
\citet{mosk03} and 24.3\% reported by \citet{mize03} for RRab 
in the bulge; the 24.3\% breaks down into 12.5\% BL1, 7.4\% BL2 and 
4.4\% BL?, with an additional 0.7\% possible PC stars. 
The incidence rate in the LMC was reported to be 
11.9\% by \citet{alco03}, which breaks down into 6.5\% BL1, 5.4\% BL2 and 
0.3\% BL?, as well as a separate 2.9\% PC incidence rate. The 
fact that our overall Blazhko incidence rate exceeds that of \citet{mize03} 
(unlikely to be simple counting statistics, considering the sample 
sizes) may be attributed to the greater number of observations in our 
data set as compared to his (four seasons instead of three) and our different 
search procedure, resulting 
in the somewhat greater sensitivity of our search. It has been suggested 
that the Blazhko effect is more prevalent in the bulge than in the LMC 
because of the different metallicities (e.g., \citealt{mosk03}). But leaving 
that question aside, what is most striking is the much higher proportion of 
BL2/BL1 stars among our classifications as compared to the previous studies.
Since we believe our detections to be robust, we suggest that this simply 
owes to our method, which greatly increased our sensitivity in the 
narrow bands associated with BL2 and PC behavior, allowing us to 
push far down into the noise to recover them.


\subsection{I-band models}
\label{sec:model}

In order to reliably extract information such as the magnitude at mean flux, 
the overall amplitude, and the behavior of the light curve during the minimum 
light interval, some degree of modeling is required. However, seeking to 
provide a faithful representation of the complex multi-periodic light curves 
exhibited by the Blazhko stars is beyond the scope of the 
present work. After consideration of many alternatives, we decided to 
adopt the simplistic approach of modeling the $I$-band light curves 
as singly-periodic Fourier sums of only six harmonics. That is, we adopted 
a model of the form
\begin{equation}
f(t) = f_0 + \sum_{k=1}^{6} [A_k \cos(2\pi k t/P) + B_k \sin(2\pi k t/P)]
\end{equation}
where $P$ is the period and $f$ indicates that all fitting was performed in 
terms of fluxes. Thus the magnitude $I_{\mathrm{mf}}$ at mean flux corresponding to 
$f_0$ can be extracted directly from the light curve. The values of 
$I_{\mathrm{mf}}$ in Table~1 have been transformed to the standard photometric 
system assuming that the intrinsic mean $V-I$ color of RR Lyrae stars 
is 0.45 (reddened according to S04); this was necessary because we 
do not have good measurements of the average $V$-band magnitude of many 
stars. The amplitudes given 
in Table~1 represent the minimum-to-maximum range (in mag) of the model.

We chose six harmonics 
because this is enough to capture the essential shape of the light curve, 
but few enough to avoid over-fitting even of particularly badly behaved 
stars. Since some light curves contain individual (or small numbers of) 
highly discrepant points that can affect the fits, we implemented a 
procedure to trim outlying points that deviated by more than 
$5\alpha_{\chi}\sigma_i$. Here $\alpha_{\chi}$ is the rms value of 
$(f_i-m_i)/\sigma_i$, 
where $f_i$, $m_i$ and $\sigma_i$ are the observed flux, model flux and 
photometric error for data point $i$. All fits were inspected to ensure that 
they had converged to visually reasonable approximations of the light curves.

\begin{figure}
\epsscale{1.0}
\plotone{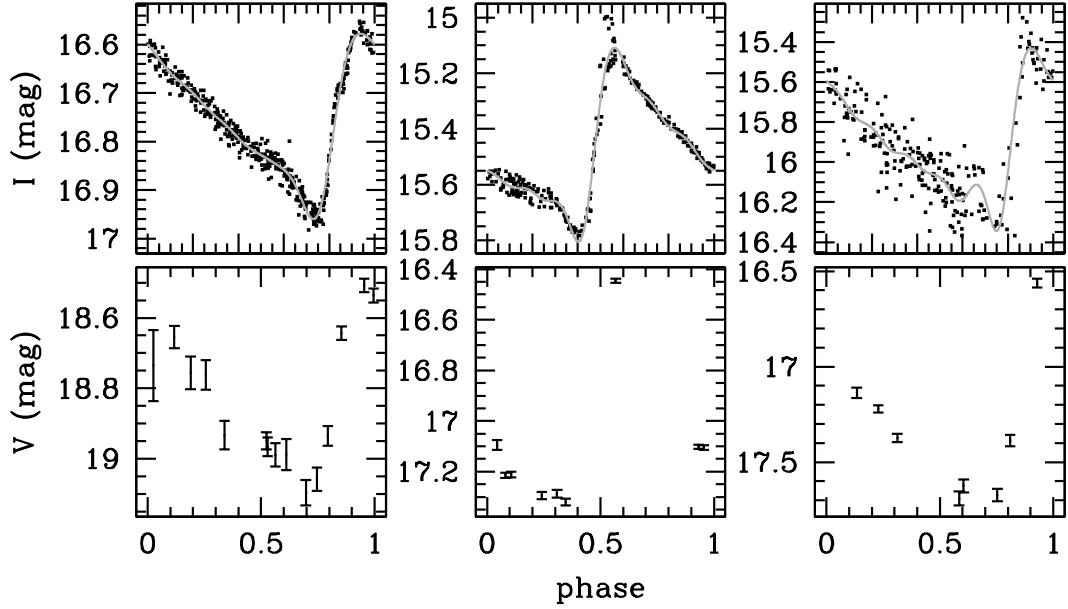}
\figcaption{
Example $V$ and $I$ light curves, with 6-harmonic Fourier model $I$ light 
curves shown. 
From left to right: sc4-579960, a plain vanilla RRab star; sc34-208801, a 
Blazhko star; sc36-709901, an RRab star showing significant scatter in the 
light curve of an unknown nature. Error bars have been omitted from the upper 
panels for the sake of clarity, and are typically at the level of 0.01--0.02.
\label{fig:lc}}
\end{figure}

As can be readily seen from the fact that most of the $\chi^2/\nu$ 
(where $\nu$ is the number of degrees of freedom) 
values presented in Table~1 are much larger than unity, 
the six-harmonic Fourier models do not provide formally acceptable fits to 
the light curves. In some cases it is possible to achieve $\chi^2/\nu\approx 1$ 
fits by increasing the number of harmonics; however, such a procedure needs to 
be tightly supervised to avoid unreasonable behavior. Moreover, 
a significant fraction of the stars simply cannot be represented 
by a model with a single, constant frequency (and its harmonics), 
so we do not include the results of higher-harmonic fits in this paper. 
Many of the stars that are intractable to single frequency models are 
Blazhko stars or PC stars, but some are not; 
detailed investigation of these cases is beyond the scope of the present paper.
Example light curves and $I$-band models are shown in Figure~\ref{fig:lc}.

\begin{figure}
\epsscale{1.0}
\plotone{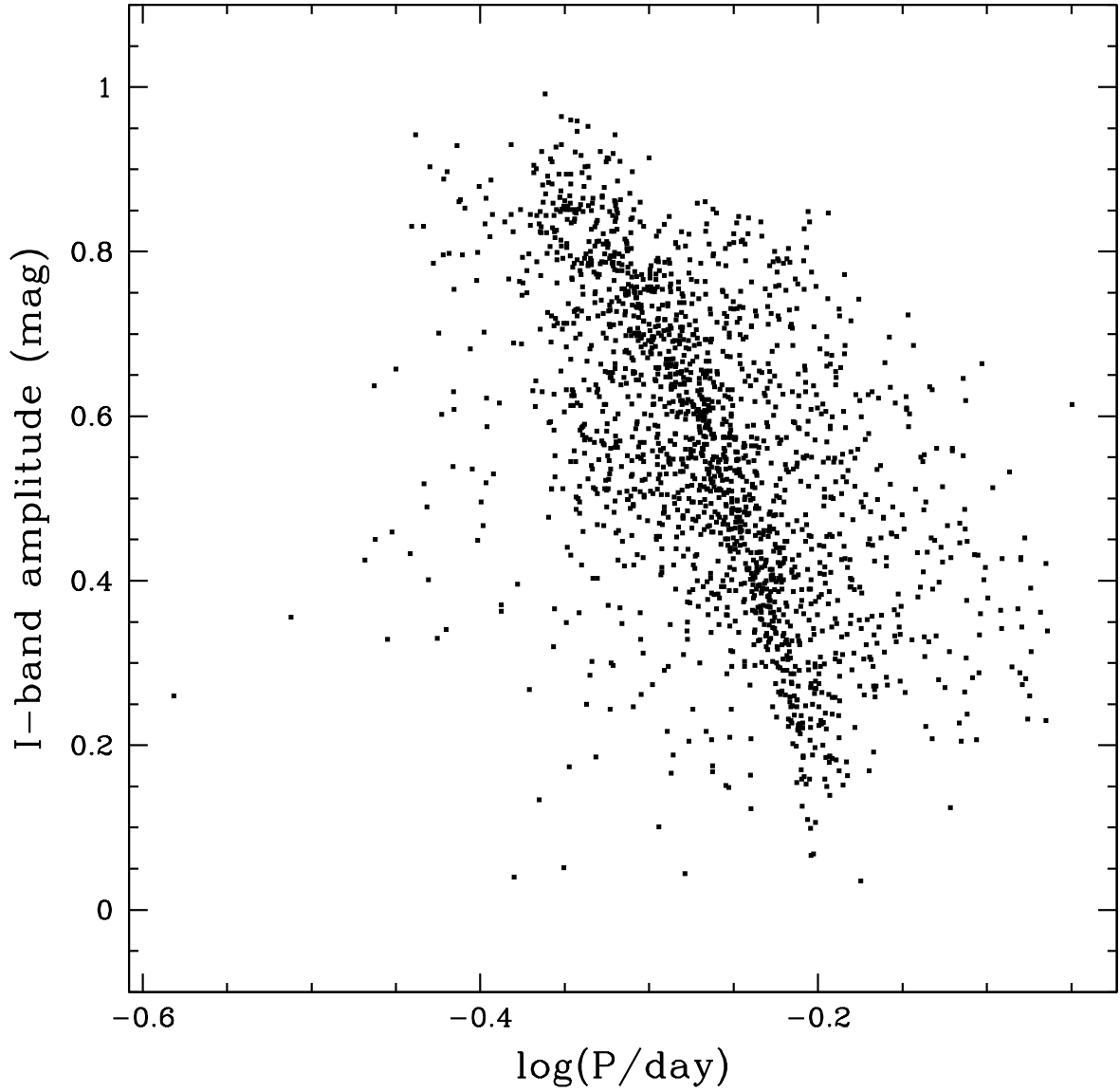}
\figcaption{
Bailey diagram for our RRab sample. As discussed in \S\ref{sec:model}, 
nine stars in our sample fall above the vertical scale ($>1.1$~mag), 
likely because of 
blending-related calibration errors. The stars with the lowest amplitudes 
($<0.1$~mag) are also probably blends.
\label{fig:peramp}}
\end{figure}

Figure~\ref{fig:peramp} 
shows the relationship between $I$-band amplitude and period, also 
known as the Bailey diagram, displaying the expected anti-correlation between 
the two quantities. Visual comparison of Figure~\ref{fig:peramp} with 
Figure~8 of \citet{sosz03}, who presented the OGLE catalog of RR Lyrae stars 
in the Large Magellanic Cloud, reveals no striking differences between their 
RRab sample and ours, other than the relative numbers. 
Such comparison also demonstrates that the contamination of our sample 
by non-fundamental mode RR Lyrae stars is negligible. Nine stars from our 
sample are not shown in Figure~\ref{fig:peramp} because they have anomalously large 
amplitudes in the range 1.1--2.8. These stars are unexpectedly 
faint, with an average $I$ magnitude of 17.5 as compared with the sample 
average of 16.2, and six of nine have 
$\Delta I_{Up}>0.1$. We suspect these are cases in which the baseline DIA 
magnitude has been set erroneously faint, possibly because of blending
(i.e., flux from the variable star has been mistakenly assigned to a 
nearby blended star), resulting in implausibly large variability 
amplitudes. In any case we exclude these stars from the analysis in 
\S\ref{sec:vmi}. We also comment that among the seven stars with 
variability amplitudes below 0.1~mag, the average $I$ magnitude of 14.3 
is rather bright; some or all of these are likely to be unresolved blends, 
leading to the bright apparent magnitudes and low variability amplitudes.
We address this and other confusion effects in the next section.


\section{($V-I$) at minimum light}
\label{sec:vmi}

The minimum light interval is usually defined to be phases of 0.5--0.8 
after maximum light (e.g., \citealt{guld05}). Here the phase has been 
normalized by $1/2\pi$ and ranges from 0--1. Since we have typically only 
5--15 $V$-band measurements per star taken at random phases, some stars in our 
catalog do not have $V$-band measurements in this interval. A further 
complication is that OGLE observations in $V$ and $I$ are by no means 
simultaneous. Since there is in general no data point in $I$ sufficiently 
close in time to a given data point in $V$, we use the light curve models 
discussed in the previous section to extract estimates of the $I$-band 
magnitude at the time corresponding to a given $V$-band observation. We 
estimate the error on the $V-I$ measurements so obtained to be 
\begin{equation}
\sigma^2_{V-I} = \sigma^2_V + {{\sum_{i=1}^m (I_{obs}-I_{model})^2}\over{m(1-\nu/N)}},
\end{equation} 
where the sum over $i$ includes the $m$ observations in $I$ within a phase interval 
of $\pm 0.05$ around the $V$-band measurement, $\nu$ is the number of degrees 
of freedom in the fit, and $N$ is the total number of $I$-band measurements.
For stars that have several $V$ data point in the minimum light interval, 
we combine them to obtain a single best estimate of 
$(V-I)_{ml}=[\sum (V-I)/\sigma^2_{V-I}]/\sum 1/\sigma^2_{V-I}$ with its 
accordant error.

From the ensemble of $V-I$ data points that fall in the minimum light interval, 
it is possible to discern a slight dependence on phase, in the sense that the 
color becomes slightly bluer with increasing phase; however this effect is weak
(total change in $V-I$ at the level of 0.01 through the 0.5--0.8 phase 
interval) and we neglect it in what follows. 


\subsection{Extracting a clean subsample}
\label{sec:clean}

Taken at face value, the $V-I$ data points obtained as described above 
include a significant 
proportion of measurements that are inaccurate for one reason or another, such 
as blending. We outline below the series of cuts we apply to go from our 
full sample of 1913 stars to the final data set of 1106 stars that yield useful 
measurements of the $V-I$ color at minimum light. These cuts are presented in 
succession, so the number of stars rejected by the cuts as given below 
corresponds to the number of additional stars rejected after the preceding 
cuts have already been applied.

\begin{figure}
\epsscale{1.0}
\plotone{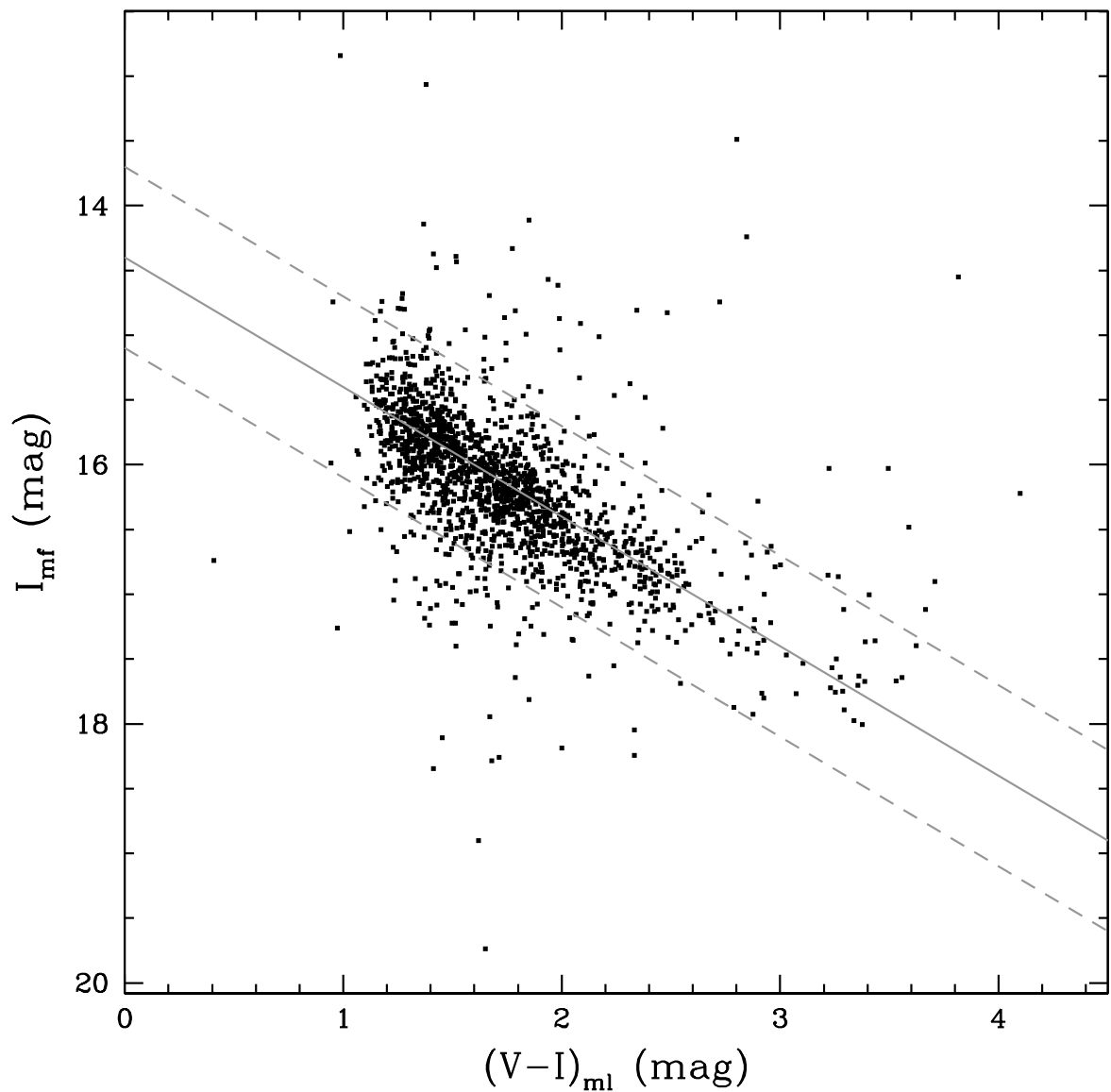}
\figcaption{
Observed color magnitude diagram for RR Lyrae stars in terms of the minimum 
light color $(V-I)_{ml}$ and the magnitude at mean flux $I_{mf}$. The three 
over-plotted lines have slopes equal to unity (consistent with the range of 
measured reddening slopes). Stars falling outside the region 
between the dashed lines are excluded from the analysis in \S\ref{sec:vmi} as 
described in \S\ref{sec:clean}.
\label{fig:cmd}}
\end{figure}

\begin{enumerate}

\item{214 stars have no $V$ observation in the minimum light interval, defined 
to be phases between 0.5--0.8 with respect to maximum light.}

\item{Seven stars have extinctions from S04 above the reliability 
threshold of $A_I>2.9$.}

\item{258 stars are predicted to have $V$ magnitudes at minimum light that 
are below the detection limit of $V_{lim}\approx 20.5$. The $V_{ml}$ magnitudes 
are predicted from the $I$-band light curve by adding the assumed 
intrinsic, minimum light color $(V-I)_{0, ml} = 0.58$, 
reddened according to S04.}

\item{62 stars have $I_{\mathrm{mf}}<13.7+(V-I)_{ml}$, that is, 
they are more than 0.7~mag brighter than the approximate reddening line 
(of slope unity, within the range of observed values from S04) 
$I_{\mathrm{mf}} = 14.4+(V-I)_{ml}$ that characterizes the bulk of the sample. 
The observed color-magnitude diagram for RR Lyrae stars is shown in 
Figure~\ref{fig:cmd}. 
The 62 stars excluded by this cut may include foreground stars, 
which may or may not be subject to the 
same amount of extinction as stars in the bulge, and stars with significant 
blending.}

\item{53 stars have $I_{\mathrm{mf}}>15.1+(V-I)_{ml}$, that is, 
they are more than 0.7~mag fainter than the ``sample'' reddening line defined 
above. These may include background stars, which could be subject to more 
extinction than stars in the bulge, and stars with zero-point errors as 
discussed in \S\ref{sec:model}.}

\item{44 stars have $|\Delta I_{Up}|\geq 0.1$, indicating that they may have 
unreliable $I$-band photometry.}

\item{56 stars have $|\Delta V_{Up}|\geq 0.1$, indicating that they may have 
unreliable $V$-band photometry.}

\item{107 stars have $\sigma_{V-I,ml}\geq 0.1$, indicating that they are 
close to the detection limit in $V$ or that the quality of the fit to the 
$I$-band light curve is especially poor.}

\item{Six stars have $(V-I)_{0,ml}$ more than $5\sigma$ away from the sample 
average (after application of all preceding cuts), where $\sigma=0.079$~mag 
is the rms deviation around the mean.}

\end{enumerate}

The resulting sample of 1106 stars has 
$\langle(V-I)_{0,ml}\rangle = 0.528\pm 0.002$. One can immediately notice that 
this value and the standard deviation quoted above 
differ significantly from the results of \citet{guld05}, who 
found $(V-I)_{0,ml,av} = 0.58\pm 0.02$ with an rms scatter of $\sigma=0.024$; 
this discrepancy is discussed in \S\ref{sec:disc}.


\subsection{Dependence on other properties}

We searched for correlations between $(V-I)_{0,ml}$ (dereddened according to 
S04) with other basic parameters: period, amplitude, and the Fourier 
phase differences $\phi_{21}\equiv \phi_2 - 2\phi_1$ and 
$\phi_{31}\equiv \phi_3 - 3\phi_1$. The latter two quantities were included 
because it has been suggested that in combination with period, these 
phases can be used to determine metallicity. Following the original 
proposal by \citet{kova95}, \citet{jurc96} 
presented a linear relationship between metallicity, period, and $\phi_{31}$ 
determined from the $V$-band light curve; recently \citet{smol05} presented 
a similar relationship calibrated in the $I$-band. 

We found statistical and visually significant relationships between 
$(V-I)_{0,ml}$ and all of the other basic parameters.
However, the interpretation of these relationships is ambiguous since 
to varying degrees period, amplitude, $\phi_{21}$ and $\phi_{31}$ all 
correlate (or anti-correlate) with one another.
In order to disentangle these dependences, we performed the following 
simple test. We searched for linear relationships between $(V-I)_{0,ml}$ and 
period (or $\log P$) and corrected the $(V-I)_{0,ml}$ values to a 
fiducial period. We then searched for correlations between these corrected 
$(V-I)_{0,ml,cor}$ values and amplitude, $\phi_{21}$ and $\phi_{31}$. 

\begin{figure}
\epsscale{1.0}
\plotone{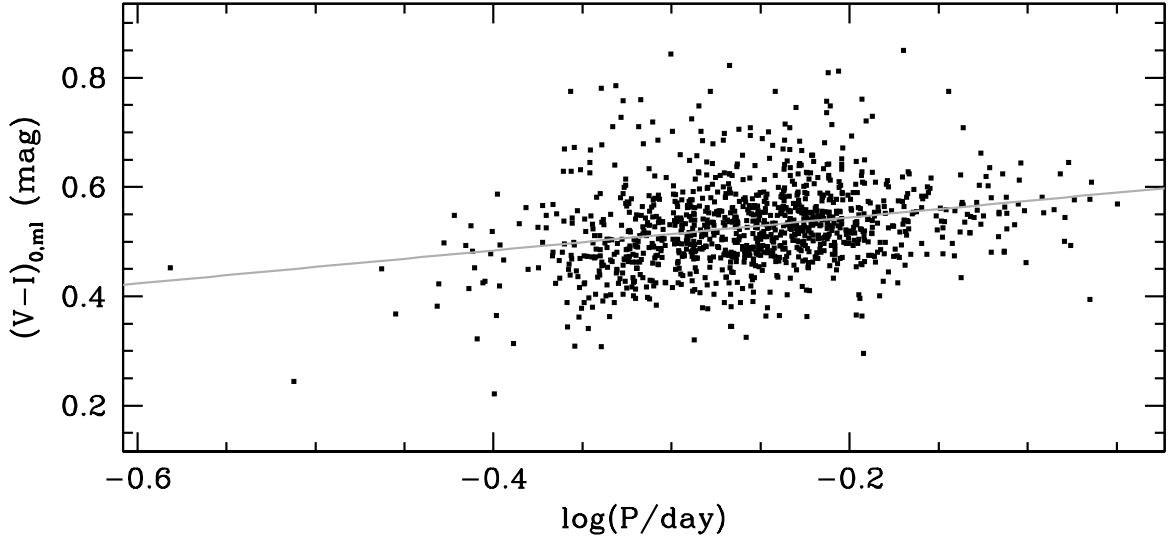}
\figcaption{
The correlation between $\log P$ and dereddened, minimum light $V-I$ color. Error 
bars have been omitted for clarity, but are typically at the level of 0.05 in 
$V-I$ and negligible in $\log P$. The fit line shown was determined from an 
unweighted linear regression and does not depend 
sensitively on the inclusion of the two points with $\log P<-0.5$.
\label{fig:lgp_vmi}}
\end{figure}

As it turns out, unweighted linear least-squares regression produces a 
relationship between $(V-I)_{0,ml}$ and $\log P$, 
\begin{equation}
\label{eqn:vmi_lgp}
(V-I)_{0,ml} = (0.525\pm 0.004) + (0.30\pm 0.06) (\log P+0.263),
\end{equation}
that is sufficient to account for virtually all of the correlations between 
$(V-I)_{0,ml}$ and the other quantities. This relationship is shown in 
Figure~\ref{fig:lgp_vmi}. After correcting $(V-I)_{0,ml}$ 
to the fiducial period of 0.546~day ($\log P = -0.263$), we compute 
Spearman's rank-order correlation coefficients $r_s$ for the remaining 
relationships between $(V-I)_{0,ml,cor}$ and amplitude, $\phi_{21}$ and 
$\phi_{31}$. The results are $r_s = -0.063$, 0.041 and 0.048 respectively, with 
corresponding null hypothesis probabilities (NHPs) of 3.6\%, 17.6\% and 11.3\%. 
For comparison, the original correlations between $(V-I)_{0,ml}$ and 
the other quantities all have NHPs $\ll 0.001$\%, meaning that they are all 
significant at much greater than 99.999\% 
probability. While it may be that weak relationships remain after taking 
out the dependence on period, 
at such low significances, it practically goes without saying that any 
remaining correlations between the quantities are invisible to the eye. 

If one accepts that RR Lyrae metallicities can be determined from the 
light curves, the fact that the relationships between $(V-I)_{0,ml}$ and 
other observables can be reduced to a simple relationship with period suggests 
that any direct dependence of $(V-I)_{0,ml}$ on metallicity should be very 
weak. This is consistent with the result of \citet{guld05} who found 
at most a weak dependence of $(V-I)_{0,ml}$ on metallicity. 

We also checked for any systematic difference of $(V-I)_{0,ml}$ between 
singly-periodic RRab and Blazhko stars. The 294 Blazhko stars that yield 
usable measurements of $(V-I)_{0,ml}$ prefer a slightly bluer color than 
the full sample: 0.519 as compared with 0.528, whereas there is no significant 
difference in the mean periods. However, the scatter in 
$(V-I)_{0,ml}$ among both Blazhko and non-Blazhko stars is much greater 
than the difference in the mean values, and in any case the 
difference is small, and the inclusion 
of the Blazhko stars does not greatly affect the overall sample mean. 


\subsection{Field-to-field variations}

\begin{figure}
\epsscale{1.0}
\plotone{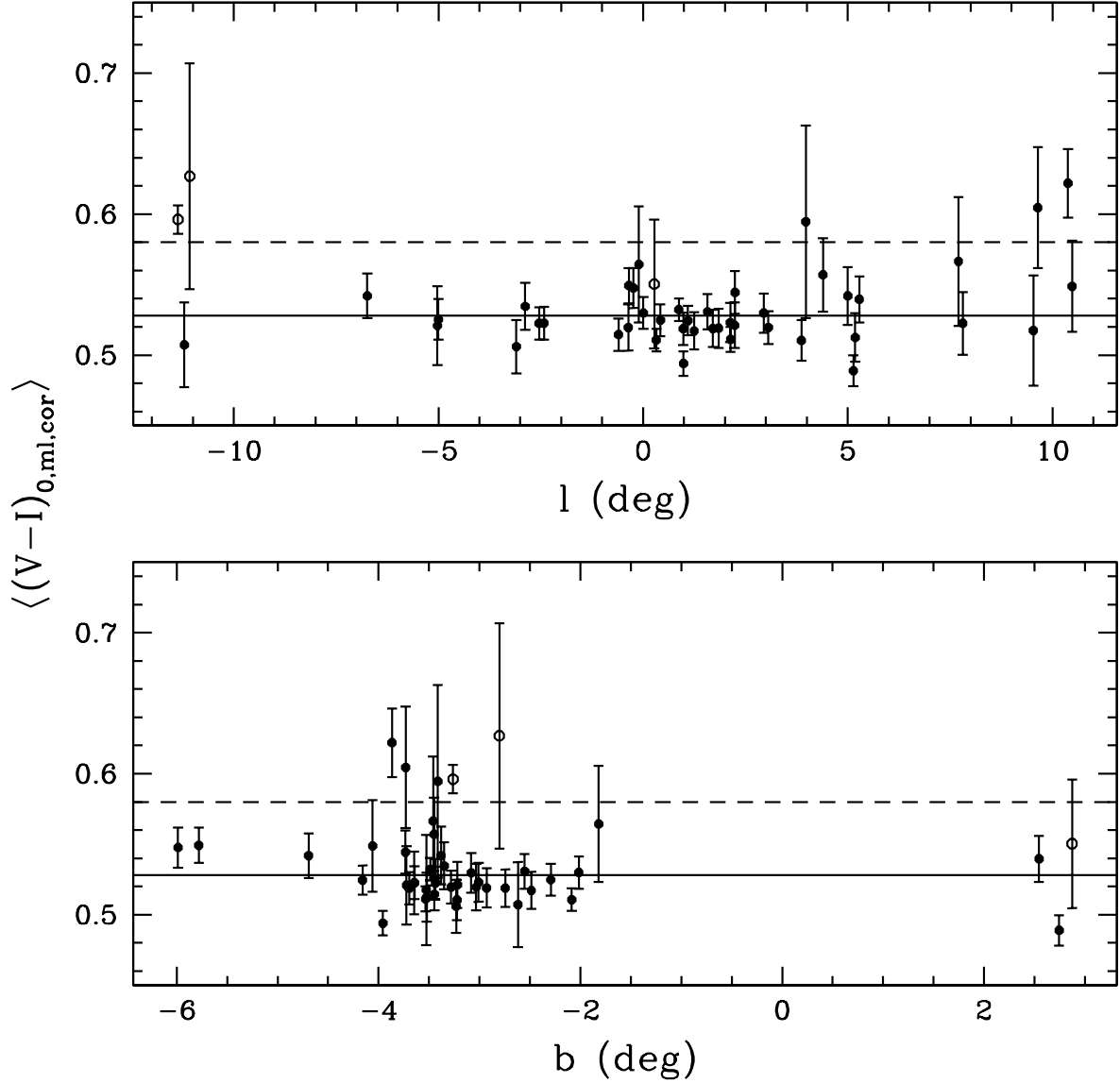}
\figcaption{
Mean $(V-I)_{0,ml,cor}$ colors versus Galactic coordinates on a field-by-field 
basis. Open circles indicate that the corresponding field has less 
than 5 RR Lyrae which yielded usable color measurements. The solid horizontal 
line indicates the sample mean of 0.528; the dashed horizontal line at 0.58 
indicates the mean color of local RR Lyrae stars from \citet{guld05}.
\label{fig:vmi_lb}}
\end{figure}

\begin{figure}
\epsscale{1.0}
\plotone{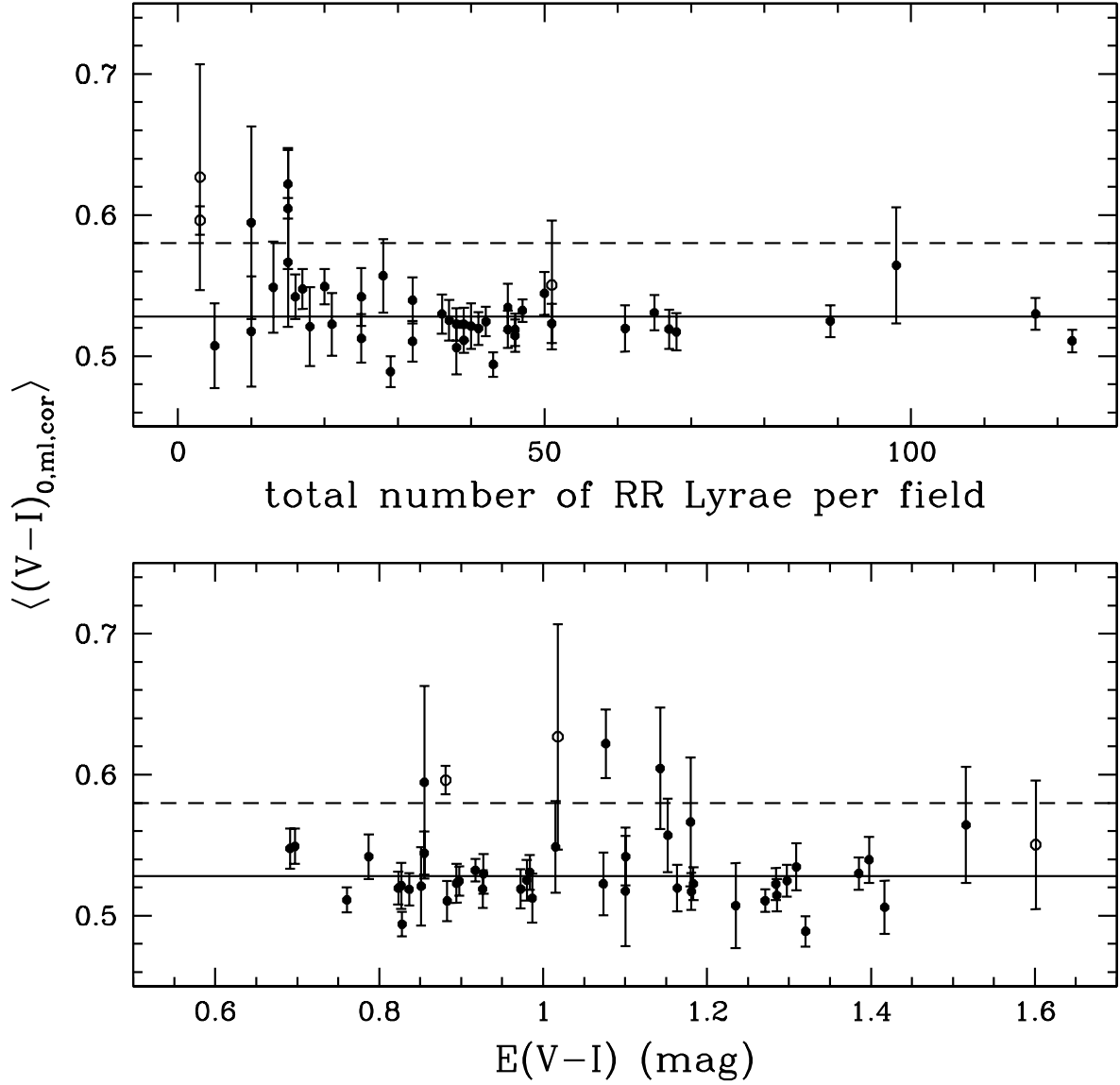}
\figcaption{
Mean $(V-I)_{0,ml,cor}$ colors versus the total number of RR Lyrae (top panel) 
and the average reddening (bottom panel) 
on a field-by-field basis. The symbolic scheme and the horizontal lines are 
identical to those in Figure~\ref{fig:vmi_lb}.
\label{fig:n.evi.vmi}}
\end{figure}

Based on the analysis of $V-I$ colors obtained from the photometry of U02 for 
RR Lyrae stars, S04 found tentative evidence that the zero-point of his 
reddening maps systematically differed for fields with Galactic longitudes $l$ 
significantly away from zero; the sense of the effect was that 
extinction corrected RR Lyrae colors appeared redder by $\sim 0.1$~mag for 
fields with large $|l|$ (see his Figure~12). In Figure~\ref{fig:vmi_lb} we 
can also see a hint of redder colors at the extremes in Galactic longitude; 
though the scatter of our points is somewhat smaller 
than S04's, the trend is no more clear in that the average colors in fields with 
large $|l|$ themselves display a large scatter. This is partly because the 
numbers of RR Lyrae stars are relatively low in these fields. Indeed, 
as the upper panel of Figure~\ref{fig:n.evi.vmi} shows, 
the redder $\langle (V-I)_{0,ml,cor}\rangle$ colors are concentrated in fields 
with relatively low stellar densities, as traced by the RR Lyrae population. 
The data from Figures~\ref{fig:vmi_lb} 
and \ref{fig:n.evi.vmi} are given in Table~4.


\section{Discussion}
\label{sec:disc}

\subsection{Relative zero points and discrepancy with field RRab colors}
\label{sec:disc_color}

In addition to the apparent dependence of the reddening zero point on 
environment, the discrepancy of 0.05~mag between the mean value of $(V-I)_{0,ml}$ 
for RRab stars in our sample and the field RRab value obtained by 
\citet{guld05} has interesting implications. Since the reddening map of S04 
was obtained by assuming that the intrinsic color of the RC in the bulge 
is the same as its local color, the 0.05~mag color offset actually represents 
a discrepancy between the RR Lyrae-to-RC color differential in the bulge 
as compared to its local value. Caution must be employed in 
trying to interpret this discrepancy, however, because systematic errors in the 
OGLE photometry may appear at approximately this level (A.\ Udalski 2005, 
private communication). While the possibility of systematic errors in the 
photometry complicates the interpretation of this discrepancy, 
what would be needed to remove it entirely would be color-dependent 
systematic errors, which are somewhat more difficult to arrange.

One possible source of color-dependent systematic errors 
is the non-standard red wing of the OGLE $I$-band filter.
The linear transformation presented by U02 to convert between OGLE 
magnitudes and Johnson-Kron-Cousins magnitudes becomes increasingly inadequate 
for redder stars. U02 calculated that this error should 
not exceed about 0.03~mag for stars with $V-I<2$, but it is non-linear 
in its dependence on color and may reach almost 0.2~mag for $V-I\approx 4$. 
The sign of the effect is that, for red stars, the $V-I$ 
colors derived from OGLE observations (after the photometric transformation 
has been applied) will be redder than the true colors, and the $I$-band 
magnitudes will be correspondingly too bright. Thus for heavily 
reddened fields, the reddening as measured by S04 may be systematically too 
large and the dereddened RR Lyrae colors artificially blue. 
However there are several problems with this explanation. The first is that the 
color offset between RC and RR Lyrae stars introduced by this phenomenon 
is a first-order correction to an already small effect, accounting for at 
most about 0.02~mag of the discrepancy for $E(V-I)<1$ (judging from Figure~2 
of U02). On top of this, the calculations of U02 neglect atmospheric 
absorption, which as they note would mitigate the errors to some 
degree. Another problem is that, since the systematic errors resulting 
from the red wing increase non-linearly for progressively redder stars, we 
would expect the color discrepancy to grow with increasing reddening, whereas 
there is no signature of this in the lower panel of Figure~\ref{fig:n.evi.vmi}.
Thus we conclude that systematic errors caused by the red wing are 
unlikely to be responsible for the observed color discrepancy.

A second possible source of systematic errors is unresolved stellar 
blends in the denser fields. The fact that the $(V-I)_{0,ml}$ colors of RRab 
stars in the less dense fields are closer to the local colors argues that 
possible problems related to crowding should be considered. However, the 
pixel-level simulations of \citet{sumi06} cast doubt on whether blending 
is a significant effect for stars as bright as RR Lyrae and RC giants 
(see their Figure~6), though it may be responsible for a relatively small 
number of magnitude outliers. A bigger problem with the blending hypothesis 
is that RR Lyrae and RC stars have comparable brightnesses, and therefore the 
populations of stars available to blend with them are practically 
indistinguishable. This 
means that on average blending should bring their colors closer together 
rather than pushing them farther apart. This is the wrong sign to explain the 
observed color discrepancy, so blending cannot be the source of the effect.

One additional factor that influences the color discrepancy is the overall 
zero point calibration of S04's extinction map. S04 used independent 
measurements of $A_V$ for 20 RRab stars in Baade's Window to determine that 
on average his RC-color method slightly overestimated the extinction, 
and consequently he applied small corrections to his reddening and 
extinction values. This correction amounts to a reduction in $E(V-I)$ of 
only 0.028, but if it had not been applied we would have observed a larger 
color discrepancy, and hence it is relevant.

Since we have eliminated what seem like the most plausible artificial causes for 
the RR Lyrae/RC color discrepancy, the most probable conclusion is that 
it is real. Such a population difference, amounting to 0.05-0.08~mag 
in $V-I$ between the bulge and the local region of the Galaxy, would not 
be very surprising. Unfortunately it is difficult to determine whether the 
effect is confined to the RC or to the RR Lyrae, or whether both classes 
of star are subject to population effects. At present there exists only 
marginal evidence for a dependence of $(V-I)_{ml}$ on metallicity for the 
RR Lyrae \citep{guld05}. While the tentative trend has the correct sign 
to explain part of the color discrepancy (assuming the bulge RR Lyrae to be 
more metal-rich than the local ones), its slope is much too 
shallow to account for all of it. Any difference between the average period 
of the field RRab stars and the average period of our sample is far too small 
to account for the effect through equation~\ref{eqn:vmi_lgp}. It may be that 
population effects are more important for the RC than for the RR Lyrae, but 
further studies of local RR Lyrae colors are needed before this can become a 
strong statement. In any case the color inconsistency indicates that 
the zero-point calibration of S04's reddening map should be treated as
uncertain at the level of 0.05 in $E(V-I)$. 

Another discrepancy between the minimum light colors of bulge 
and field RRab stars is in the star-to-star scatter. \citet{guld05} 
quote an rms scatter of only 0.024~mag, whereas the standard deviation measured 
from our sample is 0.079~mag. For comparison, a typical value of 
the photometric uncertainty $\sigma_{V-I}$ in the minimum light interval 
for our sample is 0.045~mag. Accounting for the 
observed dependence on period does not significantly reduce the scatter 
of our observed colors, nor is the scatter within most individual fields 
(0.07~mag on average for fields with more than five RRab stars) 
significantly less than in the overall sample. This star-to-star 
variation may reflect low-level random errors in the reddening map or 
errors resulting from unresolved structure in the extinction, or it may 
be that the intrinsic scatter in $(V-I)_{ml}$ has been underestimated because 
of the small number of field RRab studied.

\subsection{RC and RR Lyrae magnitudes in Baade's window}
\label{sec:disc_mag}

Another problem noted by S04 is that the extinction corrected $I$-band
magnitude of the RC in Baade's window (BW) is 14.6; the BW data were chosen 
merely to illustrate the point and do not conflict with adjacent fields. 
The expected apparent magnitude is 14.3, assuming that 
the absolute magnitude is $M_{I_{0,RC}}=-0.26\pm 0.03$ \citep{alve02} 
(from {\it Hipparcos}) and if the Galactic center (GC) distance modulus 
is $14.5\pm0.1$ (corresponding to 7.94~kpc, as given by \citealt{eise03}; 
the BW distance modulus differs from that of the GC by only about $-0.02$~mag 
according to \citealt{pacz98}). 
Since the $V$-band magnitude of RR Lyrae stars is a reasonably good
standard candle, we can check for a similar effect.
For the 158 RR Lyrae in BW that satisfy $A_I\leq 2.9$, $V_{Udal}<20.5$, 
$|\Delta V_{Up}|<0.1$, and 
$13.6+2(V-I)_{ml}\leq V_{Udal} \leq 15.1+2(V-I)_{ml}$ (similar to the 
requirements from \S\ref{sec:clean}, but adapted to the $V$-band; 
for stars that lack $V$ observations in the minimum light interval, 
we assume $(V-I)_{0,ml}=0.58$, reddened according to S04's map) we obtain 
$\langle V_{0,Udal}\rangle = 15.46\pm 0.04$, where the subscripts ``Udal'' 
and ``0'' respectively indicate that we are using mean magnitudes as 
presented by U02, corrected for extinction according to S04. 
The field-to-field scatter of $\langle V_{0,Udal}\rangle$ within BW is 
more like $\pm 0.1$~mag, so we adopt this as the approximate level of accuracy. 
We subtract a small correction of $0.03$~mag to account for the fact that 
these are mean magnitudes rather than magnitudes at mean flux; this correction 
was determined based on the $I$-band light curves and assuming a typical 
$V$-band amplitude of about one~mag. We note that since the zero point of 
S04's extinction map is fixed according to absolute extinction rather than 
color, a reddening zero point correction as suggested in \S\ref{sec:disc_color} 
would not automatically affect the derived values of extinction. Thus we obtain 
$\langle V_{0,mf}\rangle = 15.43\pm 0.1$ for the BW RR Lyrae stars.

Though the $I$-band luminosities of RR Lyrae stars are not as well calibrated, 
the OGLE $I$-band data have many advantages compared to the $V$-band. For 
the 158 stars that pass requirements 2, 4, 5 and 6 from \S\ref{sec:clean} 
(differing by only 5 stars from the 158 stars used to compute 
$\langle V_{0,Udal}\rangle$), an unweighted linear least-squares regression yields
\begin{equation}
I_{0,mf} = (-1.6\pm 0.3)(\log P + 0.263) + (14.99\pm 0.02).
\end{equation}
The mean extinction corrected $I$-band magnitude is 
$\langle I_{0,mf}\rangle=14.97\pm 0.02$ (statistical).
The slope of the relation is consistent with the value of $-1.62$ 
reported by \citet{sosz03} for the LMC RR Lyrae. However the current lack of 
a good calibration of $M_I$, especially its metallicity dependence, 
precludes our relying on the $I$-band data in what follows.

If we adopt the RR Lyrae absolute magnitude of $M_V=0.59\pm 0.03$ at 
$[Fe/H]=-1.5$ as compiled by \citet{cacc03} and assume a metallicity 
dependence of $M_V=(0.25\pm 0.05)[Fe/H] + constant$ (consistent with the 
range of slopes found in the literature), then using the average 
metallicity of $[Fe/H]=-1$ from \citet{walk91}, 
we obtain $M_V = 0.72\pm 0.04$ for RR Lyrae stars in BW. This predicts 
$\langle V_{0,mf}\rangle = 15.2\pm 0.1$ for BW RR Lyrae; in other words with this 
RR Lyrae calibration we obtain a similar discrepancy for RR Lyrae stars 
to that which S04 obtained for RC stars, though of somewhat lower statistical 
significance. But bearing in mind that this 
offset is seen in the $V$-band whereas S04's RC discrepancy is in the 
$I$-band, any implication for the overall extinction zero point is far from 
clear. Alternatively if we adopt the RR Lyrae statistical parallax calibration 
of $M_V=0.77\pm 0.13$ at $[Fe/H]=-1.6$ from \citet{goul98}, then using 
the same metallicity dependence we predict $M_V=15.42\pm 0.16$ for RR Lyrae 
stars in BW. This is uncomfortably close to the observed value. 
If the fainter RR Lyrae 
calibration is nearer to the truth, it would seem to indicate that the 
source of the RC magnitude discrepancy is intrinsic to the RC (such as a 
population effect); however the lingering uncertainties in the RR Lyrae 
absolute magnitude calibration prevent us from reaching a firm conclusion.

\subsection{The Galactic bar as traced by RC and RR Lyrae stars}
\label{sec:geom}

\begin{figure}
\epsscale{1.0}
\plotone{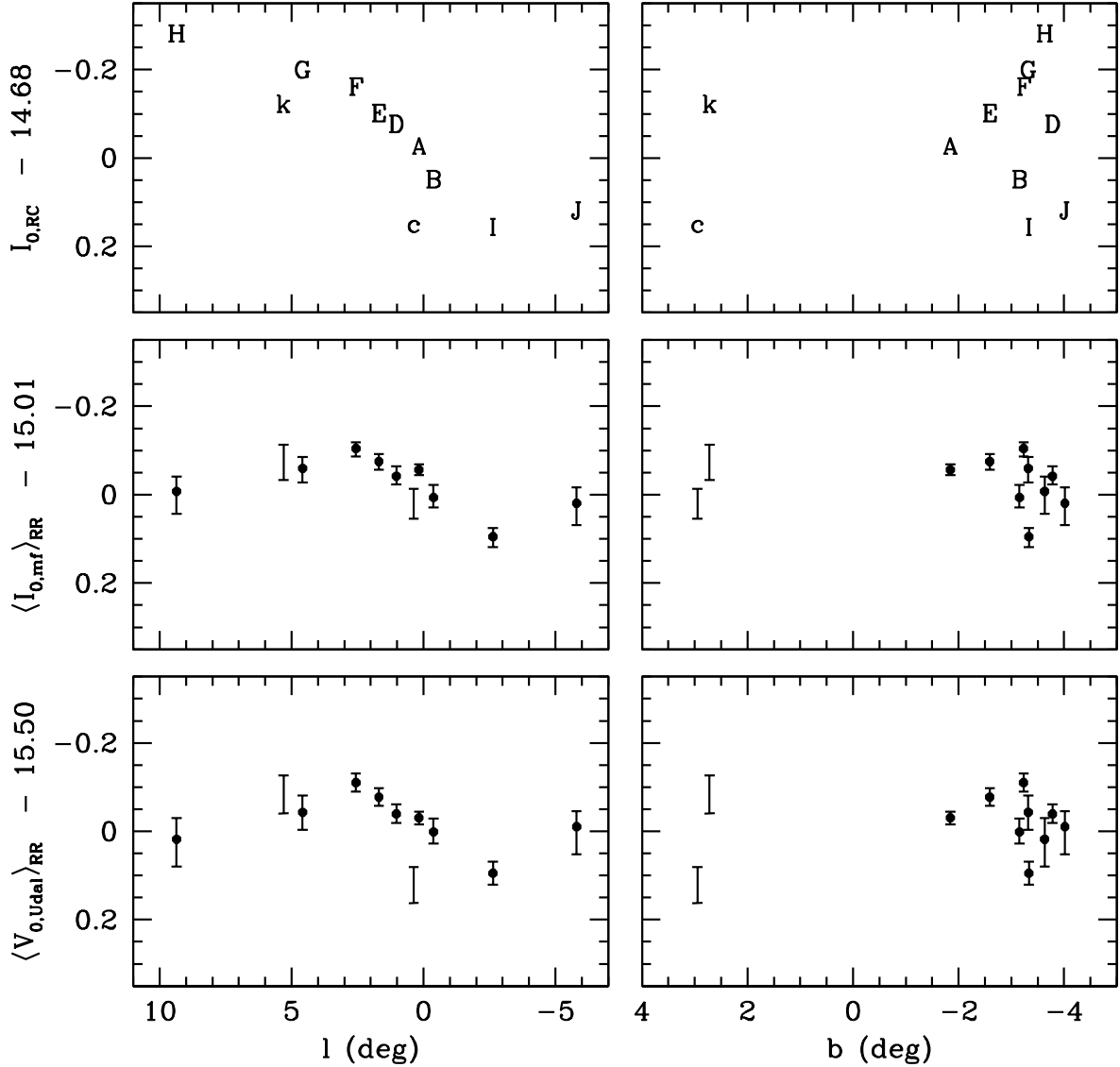}
\figcaption{
Mean extinction corrected magnitudes of RCGs 
($I$-band, top panels; from S04) and RR Lyrae stars ($I$-band, 
middle panels; $V$-band, lower panels) as functions of Galactic longitude 
(left panels) and latitude (right panels), grouped into regions A--K 
as defined by S04. The zero points approximately correspond to the mean magnitudes 
at $l=0$. The statistical errors on the points in the upper panels 
are typically less than 0.01~mag, smaller than the symbols. The regions `c' and 
`k' (shown in lower case in the upper panels and as error bars without points 
in the middle and lower panels), which are the only regions lying at positive 
Galactic latitudes, differ from the longitudinal trends evidenced by the 
remaining regions. The bar structure is clearly visible in the inner 
($|l|<3$) regions of the left-hand panels, though differences between the 
RCG and RR Lyrae magnitude offsets are apparent.
\label{fig:magbar}}
\end{figure}

While the absolute magnitude calibrations of RC and RR Lyrae stars may 
be debated, their utility as relative distance indicators (modulo population 
effects) is robust. In Figure~\ref{fig:magbar}, we show the mean 
extinction corrected magnitudes 
of RC and RR Lyrae stars as functions of Galactic coordinates, grouped into 
regions A--K following S04. In calculating the mean RR Lyrae magnitudes, 
we have applied the same requirements described in \S\ref{sec:disc_mag} for the 
$V$-band, and equivalent criteria for the $I$-band. The signature of the 
Galactic bar is clearly visible in the inner fields ($|l|<3^{\circ}$) as a trend 
from brighter to fainter apparent magnitude going from positive to negative 
Galactic longitude. The strength of this trend as reflected in the RR Lyrae 
population stands in contrast to the result of \citet{alco98a}, who found 
only weak evidence for a bar in the RR Lyrae population; however, our data 
do support the conclusion of \citet{alco98a} that the bar signature in 
the RR Lyrae population is present only in the inner fields. The fact that 
the slope of magnitude versus longitude appears somewhat shallower for the 
RR Lyrae as compared to the RC stars may reflect a contribution to 
the RR Lyrae population from the inner Galactic halo.

In addition to the signature of the bar in the inner fields, 
several other features of Figure~\ref{fig:magbar} deserve mention. 
As noticed by S04, the fields above the Galactic plane seem to differ 
from the longitudinal trends displayed by the remaining fields, though 
the significance of this effect is lower for the RR Lyrae stars. 
Also interesting is the fact that while both the 
RC and RR Lyrae magnitudes appear to recover toward the mean values 
for $l<-5^{\circ}$ (though perhaps by different amounts), the 
two populations appear to diverge for $l>4^{\circ}$, with the RR Lyrae 
magnitudes recovering toward the mean and the RC magnitudes continuing to 
brighten. The reasons for these asymmetries are not currently clear, 
but they may have important implications for the structure of the central 
Galaxy.


\section{Conclusions}
\label{sec:conc}

We have presented a catalog of 1888 fundamental mode RR Lyrae (RRab) stars 
extracted from the OGLE-II Galactic bulge data set, plus 25 double entries of 
stars detected in two fields for a total of 1913 entries. The catalog 
includes basic light curve parameters such as periods, amplitudes, mean 
magnitudes, $V-I$ measurements at minimum light, labels to indicate Blazhko 
phenomenology, and significant additional frequencies detected close to the 
main pulsation frequency, as well as extinctions derived from the map of S04 
and various additional information useful for assessing the quality of the 
photometric data for individual stars. This data set has a variety of applications 
for studies of the inner Galaxy. 

Our frequency analysis (\S\ref{sec:freq}) 
of the light curves has revealed a high incidence of the 
Blazhko phenomenon: 27.6\% of the stars show some type of clear Blazhko behavior 
and an additional 4.8\% show evidence of unstable main pulsation frequencies, 
which may be indicative of Blazhko periods longer than the time baseline 
of the observations (about four years). The Blazhko incidence rate we measure, 
which is only a lower limit, is somewhat higher than what has been found 
previously among RRab stars in the bulge, a fact that we attribute to the 
higher quality of our data set and the sensitivity of our search method. More 
strikingly, we have obtained a much higher ratio of BL2 stars (with symmetrical 
frequency triplets, i.e., one additional significant frequency on either side 
of the main pulsation frequency) to BL1 stars (with only a single additional 
significant frequency) than has been found previously. Within the limits of 
our sensitivity, we find BL2s to be about 1.7 times more common than BL1s, 
whereas previous studies of the bulge \citep{mize03} and the LMC \citep{alco03} 
found BL2/BL1 incidence ratios of 0.59 and 0.83 respectively. Since we 
believe our detections of additional frequencies to be robust, we attribute 
this large increase of the BL2/BL1 ratio to the specific and 
very sensitive method we employ to detect additional, symmetrical 
frequency components. Furthermore, we have identified several instances 
of RRab stars that have two pairs of symmetrical frequency triplets, that is, 
stars with two separate sets of BL2 sidebands. These frequency 
quintuplets are unevenly spaced. This phenomenon may be important for 
understanding the long-unsolved origin of the Blazhko effect.

From the comparison of $(V-I)_{ml}$ (minimum light) colors of bulge RR Lyrae 
stars, dereddened according to S04, with field RR Lyrae colors, we conclude that
there is a discrepancy of 0.05--0.08~mag between the RR Lyrae-to-red clump (RC) 
color differential of the bulge population (measured from OGLE data) as compared 
to the local population. The sense of the effect is that the color separation is 
greater for the bulge population. 
We evaluate likely sources of systematic color errors and 
conclude that the color discrepancy is probably real. If this is correct then the 
most likely explanation seems to be the influence of metallicity on the color 
of the RC, or possibly the RR Lyrae, although there is some evidence 
against the latter possibility. We have observed a weak dependence of 
$(V-I)_{0,ml}$ on period (redder at longer periods), but find no evidence 
of a further dependence on metallicity as estimated from the light curves. 
The trend with period cannot account for the observed color discrepancy. 
While the color discrepancy does not affect the conclusion 
that the reddening slope toward the bulge is anomalously flat, it does 
cast some doubt on the zero-point accuracy of S04's reddening map. 
If the RR Lyrae are more reliable reddening tracers, then the $E(V-I)$ 
values reported by S04 should be reduced by approximately 0.05~mag, with 
some variation from field to field as indicated in Table~4.
We additionally measure an unexpectedly 
high star-to-star scatter of $(V-I)_{0,ml}$ about 0.07~mag, which is larger 
than the photometric uncertainties and probably results from 
unresolved structure in the extinction or other random-type 
errors in the reddening map.

We exploit the approximately uniform mean $V$-band luminosities of the RRab 
stars to study the distance to and geometry of the inner Galaxy, and we 
compare these results to what S04 obtained from studying the RC. Since a 
coherent picture has not emerged, we summarize some relevant facts below.

\begin{itemize}

\item{The Galactocentric distance modulus has been measured to be $14.5\pm0.1$~mag 
by \citet{eise03} from the orbit of a star around the central black hole.}

\item{Correcting for extinction according to S04, RC stars indicate 
a distance modulus to Baade's Window (which should be virtually identical 
to the Galactocentric distance modulus) of $14.86\pm 0.04$~mag, measured 
in the $I$-band, assuming that 
they have the same luminosities as local RC stars.
Likely corrections to the reddening zero point (and concomitant corrections 
to the extinction, assuming that the measured reddening slopes hold all the 
way to $E(V-I)=0$, for which there is no direct observational evidence) 
have the wrong sign to bring the distance modulus down.}

\item{The distance modulus to Baade's Window measured from RRab stars 
is somewhere between approximately 14.4--14.8~mag, depending mainly on the 
absolute magnitude calibration adopted. This is measured in the $V$ band.}

\item{Both the RC and the RR Lyrae stars clearly reveal the signature of 
the Galactic bar in fields with Galactic longitude $|l|<3$. While the 
sense of the effect is the same (brighter at positive longitude as compared 
to negative longitude), the slopes 
are different. The magnitude difference from $l\approx -3^{\circ}$ to 
$l\approx 3^{\circ}$ is approximately 0.35--0.4~mag for the RC and 0.2--0.25~mag 
for the RR Lyrae.}

\item{Both the RC and RR Lyrae magnitudes differ by 0.1--0.2~mag 
between fields approximately symmetrically 
located above and below the Galactic plane, with 
the fields at positive latitude being fainter. This assumes there is no 
significant extinction zero-point mismatch between the fields on opposite 
sides of the Galactic plane.}

\item{There is a significant mismatch between the RC and RR Lyrae magnitude trends 
between about $3^{\circ}<l<10^{\circ}$, increasing toward larger $l$, with the
RR Lyrae becoming fainter and the RC becoming brighter; the discrepancy 
between the two trends reaches approximately 0.25~mag at $l\approx 10^{\circ}$.}

\end{itemize}

\section*{Acknowledgments}

We are very pleased to acknowledge the countless helpful suggestions, 
interesting ideas and constant support of B.\ Paczy{\'n}ski. We also 
sincerely thank the OGLE team for their continuing efforts and generosity 
in sharing data. 



\clearpage


\clearpage


\clearpage


\clearpage


\clearpage


\clearpage


\clearpage


\clearpage



\begin{deluxetable}{crccccccrccccccc} 

\tabletypesize{\tiny} 

\tablewidth{0pt} 

\setlength{\tabcolsep}{0.05in}

\tablecaption{RRab sample.}

\tablehead{
field & ID & $\alpha_{2000}$ & $\delta_{2000}$ & Period & $I_{mf}$ & Amp. & N(I) & $\chi^2/\nu$ & $V$ & $(V-I)_{ml}$ & $A_V$ & $A_I$ & label & noise & multID \\
BUL\_SC & & (hours) & (deg) & (day) & (mag) & (mag) & & & (mag) & (mag) & (mag) & (mag) & & (mmag) & \\
(1) & (2) & (3) & (4) & (5) & (6) & (7) & (8) & (9) & (10) & (11) & (12) & (13) & (14) & (15) & (16)
}

\input{tab1}

\tablecomments{Columns two, three, four and ten are from U02. Column six 
gives the magnitude at mean flux,column seven gives the minimum-to-maximum 
$I$-band amplitude, and column nine gives $\chi^2$ per degree of freedom; 
all these quantities are derived from 6-harmonic Fourier models 
(see \S\ref{sec:model}). An asterisk in column six or ten indicates that the 
photometry may be unreliable as assessed in \S\ref{sec:data}.
Column eight gives the number of $I$-band observations.
Typical errors (not shown) for the quantities in columns five, six and seven are 
approximately: less than $10^{-5}$~day, 0.005~mag and 0.02~mag. Columns~12 and 13 
are from \citet{sumi04}. Column~14 indicates the classification, and column~15 
gives the approximate noise level in the Blazhko range, as described 
in \S\ref{sec:freq}. For stars detected in two overlapping fields, column~16 
provides a unique number from 1--25 to identify the matching data set. 
(The complete version of this table will appear electronically.)}

\end{deluxetable}


\begin{deluxetable}{crccc} 
\tabletypesize{\tiny} 
\tablewidth{0pt} 
\setlength{\tabcolsep}{0.05in}
\tablecaption{RRab candidates rejected from the sample.}
\tablehead{
field & ID & $\alpha_{2000}$ & $\delta_{2000}$ & flag \\
BUL\_SC & & (hours) & (deg) & \\
(1) & (2) & (3) & (4) & (5)
}
\input{tab2}
\tablecomments{Column two gives the star catalog number from U02. 
Column five indicates why the star was rejected: if there were less than 25 
good data points in the $I$-band the flag was incremented by +1, if the light curve 
was visually inconsistent with an RRab light curve (e.g., constant) the flag 
was incremented by +2, and if the star was blended with a true RRab star the flag 
was incremented by +4.}
\end{deluxetable}


\begin{deluxetable}{crccccccccccccc} 

\rotate 

\tabletypesize{\tiny} 

\tablewidth{0pt} 

\setlength{\tabcolsep}{0.05in}

\tablecaption{Blazhko stars.}

\tablehead{
field & ID & label & noise & limit & $\Delta f_1$ & ${\cal{A}}_1$ & $\Delta f_2$ & ${\cal{A}}_2$ & $\Delta f_3$ & ${\cal{A}}_3$ & $\Delta f_4$ & ${\cal{A}}_4$ & $\Delta f_5$ & ${\cal{A}}_5$ \\
BUL\_SC & & & (mmag) & (mmag) & ($10^{-3}d^{-1}$) & (mmag) & ($10^{-3}d^{-1}$) & (mmag) & ($10^{-3}d^{-1}$) & (mmag) & ($10^{-3}d^{-1}$) & (mmag) & ($10^{-3}d^{-1}$) & (mmag) \\
(1) & (2) & (3) & (4) & (5) & (6) & (7) & (8) & (9) & (10) & (11) & (12) & (13) & (14) & (15)
}

\input{tab3}

\tablecomments{Columns one through four are reproduced from Table~1. Column 
five gives the largest amplitude observed inside the equidistant triplet 
frequency region, for the BL1 stars. Columns six through 15 list the 
significant frequencies detected within 0.1~day$^{-1}$ of the main pulsation 
frequency and their amplitudes. All frequencies are in units of 
$10^{-3}$day$^{-1}$ and all amplitudes are in milli-magnitudes. (The complete 
version of this table will appear electronically.)}

\end{deluxetable}


\begin{deluxetable}{crrrrcccc} 
\tabletypesize{\tiny} 
\tablewidth{0pt} 
\setlength{\tabcolsep}{0.05in}
\tablecaption{RRab numbers and colors on a field by field basis.}
\tablehead{
field & l & b & N(V-I) & N(RR) & $\langle (V-I)_{0,ml}\rangle$ & $\langle (V-I)_{0,ml,cor}\rangle$ & $\sigma_{cor}$ & $\langle E(V-I)\rangle$ \\
BUL\_SC & (deg) & (deg) & & & (mag) & (mag) & (mag) & (mag) \\
(1) & (2) & (3) & (4) & (5) & (6) & (7) & (8) & (9) 
}
\input{tab4}
\tablecomments{Columns four and five list, for each field, the number of RR 
Lyrae that yielded a usable measurement of $(V-I)_{ml}$ and the total number of 
RR Lyrae in our catalog. Column eight gives the standard deviation of 
$\langle (V-I)_{0,ml,cor}\rangle$.}
\end{deluxetable}



\end{document}

%% file: tab1.tex
\startdata
 1 & 190423 & 18.04199 & -30.4199 & 0.43308 &     15.73 & 0.63 & 232 &  40.1 &     16.83 & 1.29$\pm$0.14 &  1.52 &  0.75 &      BL2 &  14.7 & \nodata \\ 
 1 & 566234 & 18.05063 & -30.3314 & 0.56371 &     16.16 & 0.35 & 233 &  58.6 &     17.51 & 1.39$\pm$0.10 &  1.64 &  0.80 &    BL2+? &  17.6 & \nodata \\ 
 1 & 576932 & 18.04904 & -30.3100 & 0.68769 &     15.69 & 0.36 & 253 &   2.0 &     16.98 & 1.42$\pm$0.04 &  1.70 &  0.83 &  \nodata &   3.3 & \nodata \\ 
 1 & 587412 & 18.04701 & -30.2152 & 0.49277 &     16.10 & 0.59 & 251 &  21.4 &     17.26 & 1.34$\pm$0.08 &  1.68 &  0.83 &      BL2 &  14.8 & \nodata \\ 
 1 &  36798 & 18.03571 & -30.2014 & 0.59743 &     15.84 & 0.29 & 251 &   4.8 &     17.34 & 1.50$\pm$0.02 &  1.72 &  0.85 &      BL2 &   4.2 & \nodata \\ 
 1 & 410183 & 18.04384 & -30.2007 & 0.59381 &     15.52 & 0.36 & 252 &   2.2 &     16.75 & 1.36$\pm$0.02 &  1.58 &  0.78 &  \nodata &   3.2 & \nodata \\ 
 1 & 421285 & 18.04672 & -30.1716 & 0.51940 & 15.91$^*$ & 0.39 & 254 &  17.3 &     17.22 & 1.27$\pm$0.07 &  1.73 &  0.85 &      BL2 &   9.9 & \nodata \\ 
 1 & 597902 & 18.04754 & -30.1651 & 0.68050 &     15.81 & 0.19 & 252 &   1.2 &     17.25 & 1.49$\pm$0.03 &  1.87 &  0.92 &  \nodata &   2.7 & \nodata \\ 
 1 & 608436 & 18.04774 & -30.1342 & 0.61593 &     15.91 & 0.21 & 252 &   7.6 &     17.31 & 1.42$\pm$0.04 &  1.66 &  0.82 &      BL1 &   4.9 & \nodata \\ 
 1 & 431836 & 18.04421 & -30.1187 & 0.59450 &     15.90 & 0.34 & 242 &   6.7 &     17.25 & 1.41$\pm$0.10 &  1.70 &  0.83 &      BL1 &  12.1 & \nodata \\ 
 1 &  59340 & 18.03466 & -30.1023 & 0.50249 &     15.76 & 0.78 & 250 &   7.9 &     17.43 & 1.59$\pm$0.05 &  1.96 &  0.97 &  \nodata &  12.5 & \nodata \\ 
 1 &  59700 & 18.03554 & -30.1069 & 0.50791 &     16.30 & 0.66 & 251 &  26.9 &     17.87 & 1.62$\pm$0.03 &  2.06 &  1.01 &       PC &  22.2 & \nodata \\ 
 1 &  59431 & 18.03774 & -30.0859 & 0.49027 &     16.11 & 0.65 & 251 &  25.7 &     17.34 & 1.32$\pm$0.11 &  1.79 &  0.88 &      BL2 &  14.8 & \nodata \\ 
 1 & 619252 & 18.05103 & -30.0667 & 0.46509 &     16.96 & 0.57 & 202 &   8.6 & 18.15$^*$ & 1.38$\pm$0.10 &  1.47 &  0.72 &       PC &  24.5 & \nodata \\ 
 1 & 268899 & 18.03951 & -30.0087 & 0.47363 &     16.03 & 0.81 & 251 &   4.6 &     17.27 & 1.30$\pm$0.03 &  1.61 &  0.79 &  \nodata &   3.7 & \nodata \\ 
 1 & 452435 & 18.04316 & -30.0038 & 0.46722 &     16.01 & 0.77 & 250 &   5.2 &     17.37 & 1.26$\pm$0.04 &  1.50 &  0.73 &  \nodata &   5.5 & \nodata \\ 
 1 & 629186 & 18.04859 & -29.9993 & 0.47863 &     15.80 & 0.86 & 251 &   8.9 &     16.89 & 1.25$\pm$0.04 &  1.50 &  0.74 &  \nodata &   4.8 & \nodata \\ 
 1 & 629168 & 18.04927 & -30.0017 & 0.55032 &     15.88 & 0.48 & 250 &  17.4 &     17.02 & 1.21$\pm$0.05 &  1.48 &  0.72 &  \nodata &  19.5 & \nodata \\ 
 1 & 269030 & 18.04235 & -29.9823 & 0.49502 &     15.70 & 0.57 & 252 & 228.7 &     16.95 & 1.38$\pm$0.20 &  1.44 &  0.71 &       PC &  50.7 & \nodata \\ 
 1 & 452510 & 18.04262 & -29.9895 & 0.55213 &     15.82 & 0.53 & 252 &   3.9 &     17.02 & 1.19$\pm$0.02 &  1.38 &  0.68 &      BL1 &   5.1 & \nodata \\ 
 1 & 452200 & 18.04311 & -29.9887 & 0.82207 &     14.99 & 0.29 & 252 &   1.7 &     16.26 & 1.27$\pm$0.02 &  1.40 &  0.69 &  \nodata &   2.0 & \nodata \\ 
 1 & 452506 & 18.04626 & -29.9898 & 0.55722 &     15.56 & 0.51 & 246 &   2.8 &     16.69 & 1.23$\pm$0.03 &  1.33 &  0.65 &  \nodata &   4.0 & \nodata \\ 
 1 & 280334 & 18.03862 & -29.9486 & 0.49635 &     16.37 & 0.73 & 250 &   2.3 &     17.75 & 1.41$\pm$0.05 &  1.78 &  0.87 &  \nodata &   5.4 & \nodata \\ 
 1 &  91419 & 18.03720 & -29.9312 & 0.44857 &     16.18 & 0.86 & 251 &   6.3 &     17.36 & 1.39$\pm$0.04 &  1.72 &  0.84 &  \nodata &   7.0 & \nodata \\ 
 1 & 463376 & 18.04528 & -29.9054 & 0.44442 &     15.76 & 0.81 & 201 &   6.5 & 17.12$^*$ & 1.22$\pm$0.04 &  1.42 &  0.69 &  \nodata &   5.8 & \nodata \\ 
 1 & 291846 & 18.03947 & -29.8939 & 0.45411 &     15.82 & 0.79 & 252 &   9.3 &     16.97 & 1.29$\pm$0.04 &  1.60 &  0.78 &  \nodata &   4.5 & \nodata \\ 
 1 & 652490 & 18.05015 & -29.8977 & 0.55453 &     15.21 & 0.57 & 251 &   4.0 &     16.32 & 1.14$\pm$0.02 &  1.30 &  0.64 &       PC &   5.1 & \nodata \\ 
 1 & 474984 & 18.04581 & -29.8856 & 0.60023 &     15.69 & 0.48 & 228 &   2.2 &     16.91 & 1.32$\pm$0.02 &  1.55 &  0.76 &  \nodata &   4.4 & \nodata \\ 
 1 & 474747 & 18.04416 & -29.8656 & 0.52955 &     15.69 & 0.64 & 247 &  22.6 &     16.71 & 1.30$\pm$0.02 &  1.52 &  0.75 &      BL2 &  11.8 & \nodata \\ 
 1 & 116187 & 18.03639 & -29.8346 & 0.61174 &     15.84 & 0.30 & 247 &   1.4 &     17.07 & 1.27$\pm$0.03 &  1.49 &  0.73 &  \nodata &   3.2 & \nodata \\ 
 1 & 665317 & 18.04934 & -29.8108 & 0.42413 &     15.64 & 0.75 & 251 &  31.8 &     16.76 & 1.27$\pm$0.06 &  1.38 &  0.68 &      BL1 &  14.7 & \nodata \\ 
 1 & 130494 & 18.03560 & -29.7847 & 0.48306 &     15.88 & 0.79 & 251 &   5.2 &     17.15 & 1.31$\pm$0.04 &  1.64 &  0.80 &  \nodata &   8.2 & \nodata \\ 
 1 & 130531 & 18.03746 & -29.7788 & 0.35476 & 16.06$^*$ & 0.66 & 251 &   3.0 &     17.56 & 1.51$\pm$0.04 &  1.54 &  0.76 &  \nodata &   5.1 & \nodata \\ 
 1 & 130718 & 18.03502 & -29.7495 & 0.63271 &     15.92 & 0.42 & 249 &  12.1 &     17.35 & 1.42$\pm$0.05 &  1.43 &  0.70 &      BL? &   6.5 & \nodata \\ 
 1 & 317844 & 18.04071 & -29.7296 & 0.49322 &     15.54 & 0.71 & 251 &   5.4 &     16.92 & 1.28$\pm$0.02 &  1.39 &  0.68 &  \nodata &   4.7 & \nodata \\ 
 1 & 678325 & 18.04711 & -29.7389 & 0.42179 &     16.15 & 0.76 & 251 &   2.6 &     17.20 & 1.14$\pm$0.03 &  1.35 &  0.67 &       PC &   5.8 & \nodata \\ 
 1 & 144336 & 18.03765 & -29.7049 & 0.42434 &     15.97 & 0.78 & 250 &   6.0 &     17.26 & 1.26$\pm$0.05 &  1.49 &  0.73 &  \nodata &   6.2 & \nodata \\ 
 1 & 144492 & 18.03646 & -29.6825 & 0.40112 &     16.02 & 0.87 & 249 &   8.9 &     17.22 & 1.22$\pm$0.04 &  1.56 &  0.77 &  \nodata &   7.9 & \nodata \\ 
 1 & 144504 & 18.03719 & -29.6803 & 0.47787 &     16.14 & 0.82 &  90 &   3.7 &     17.31 & 1.22$\pm$0.04 &  1.55 &  0.76 &  \nodata &  12.2 & \nodata \\ 
 1 & 331128 & 18.04075 & -29.6781 & 0.63828 &     15.46 & 0.22 & 252 &   1.3 &     16.71 & 1.35$\pm$0.02 &  1.54 &  0.76 &  \nodata &   3.4 & \nodata \\ 
 1 & 512310 & 18.04424 & -29.6729 & 0.66002 &     15.90 & 0.56 & 242 &   1.7 &     17.31 & 1.44$\pm$0.02 &  1.89 &  0.93 &  \nodata &   3.9 & \nodata \\ 
 1 & 700895 & 18.04823 & -29.6463 & 0.45696 &     15.59 & 0.42 & 250 &   1.4 &     16.93 & 1.31$\pm$0.01 &  1.63 &  0.80 &  \nodata &   3.2 & \nodata \\ 
 1 & 169019 & 18.03426 & -29.5863 & 0.55208 &     16.30 & 0.63 & 230 &   2.3 &     17.84 & 1.47$\pm$0.06 &  2.26 &  1.11 &  \nodata &   7.0 & \nodata \\ 
 1 & 168859 & 18.03648 & -29.5699 & 0.56632 &     15.82 & 0.47 & 249 &   2.2 &     17.09 & 1.30$\pm$0.02 &  1.84 &  0.90 &  \nodata &   4.8 & \nodata \\ 
 1 & 366334 & 18.03906 & -29.5067 & 0.48692 &     16.67 & 0.62 & 247 &  27.3 &     18.64 & 2.05$\pm$0.09 &  2.78 &  1.36 &      BL2 &  26.1 & \nodata \\ 
 1 & 366819 & 18.03967 & -29.5059 & 0.53524 &     17.06 & 0.40 & 246 &   9.6 &     19.20 & 2.14$\pm$0.06 &  2.82 &  1.38 &      BL2 &  18.5 & \nodata \\ 
 2 & 425494 & 18.07648 & -29.3191 & 0.75001 &     15.42 & 0.27 & 234 &   1.1 &     16.75 & 1.31$\pm$0.02 &  1.50 &  0.74 &  \nodata &   2.4 & \nodata \\ 
 2 &    464 & 18.06687 & -29.3052 & 0.50933 &     15.40 & 0.52 & 234 &   2.7 &     16.65 & 1.18$\pm$0.02 &  1.38 &  0.68 &  \nodata &   4.4 & \nodata \\ 
 2 & 436452 & 18.07862 & -29.2820 & 0.59032 &     14.68 & 0.48 & 235 &   8.1 &     15.93 & 1.27$\pm$0.03 &  1.50 &  0.74 &  \nodata &   4.5 & \nodata \\ 
 2 & 436573 & 18.07838 & -29.2569 & 0.49512 &     15.77 & 0.51 & 233 &  68.1 &     16.96 & 1.30$\pm$0.08 &  1.64 &  0.80 &      BL2 &  20.0 & \nodata \\ 
\enddata

%% file: tab2.tex
\startdata
 3 & 507991 & 17.89574 &  -30.0205 & 2 \\ 
 3 & 536371 & 17.89554 &  -29.8880 & 2 \\ 
 4 & 454331 & 17.91266 &  -29.9422 & 2 \\ 
 4 &  59232 & 17.90324 &  -29.9089 & 4 \\ 
 4 &  93538 & 17.90302 &  -29.7675 & 3 \\ 
 4 & 108677 & 17.90510 &  -29.6869 & 3 \\ 
20 & 548861 & 17.99151 &  -28.6643 & 2 \\ 
21 & 471290 & 18.00916 &  -29.3047 & 2 \\ 
22 & 410981 & 17.94948 &  -31.1071 & 2 \\ 
22 & 176263 & 17.93831 &  -30.3981 & 3 \\ 
27 & 443264 & 17.80931 &  -35.1738 & 2 \\ 
27 & 640784 & 17.81499 &  -34.9702 & 3 \\ 
28 & 116982 & 17.78153 &  -37.4575 & 3 \\ 
28 & 327437 & 17.79315 &  -37.3613 & 3 \\ 
28 & 136056 & 17.78086 &  -37.2941 & 3 \\ 
28 & 142101 & 17.78285 &  -37.2114 & 3 \\ 
28 & 142130 & 17.78419 &  -37.2014 & 3 \\ 
28 & 346566 & 17.79030 &  -37.1777 & 3 \\ 
28 &  43959 & 17.77705 &  -37.1680 & 3 \\ 
28 & 346511 & 17.79238 &  -37.1371 & 3 \\ 
28 & 260827 & 17.78959 &  -37.0154 & 3 \\ 
28 &  64509 & 17.77878 &  -36.9898 & 3 \\ 
28 &  77184 & 17.77991 &  -36.8848 & 3 \\ 
28 & 380745 & 17.79189 &  -36.8554 & 3 \\ 
28 & 380822 & 17.79203 &  -36.8366 & 3 \\ 
30 & 634223 & 18.02836 &  -29.0187 & 2 \\ 
30 & 352519 & 18.02028 &  -28.5756 & 3 \\ 
31 & 562207 & 18.04094 &  -28.3241 & 3 \\ 
33 & 130969 & 18.08350 &  -28.7448 & 2 \\ 
34 & 298657 & 17.96759 &  -29.4475 & 3 \\ 
34 & 682061 & 17.97475 &  -28.8035 & 2 \\ 
34 & 703555 & 17.97453 &  -28.7290 & 2 \\ 
37 &  55168 & 17.86717 &  -30.0903 & 3 \\ 
37 & 405002 & 17.87823 &  -29.9921 & 2 \\ 
37 & 440744 & 17.87862 &  -29.8314 & 2 \\ 
37 & 473842 & 17.87864 &  -29.5821 & 2 \\ 
39 & 684442 & 17.93639 &  -29.7803 & 2 \\ 
40 & 479913 & 17.85459 &  -32.7850 & 3 \\ 
41 & 316456 & 17.87206 &  -33.5204 & 3 \\ 
45 & 356751 & 18.06310 &  -30.4038 & 2 \\ 
45 & 245930 & 18.05976 &  -30.0334 & 1 \\ 
45 & 245283 & 18.05976 &  -30.0339 & 1 \\ 
45 & 254862 & 18.05998 &  -30.0311 & 3 \\ 
\enddata

%% file: tab3.tex
\startdata
 1 & 190423 &      BL2 & 14.7 & \nodata &  30.4 &  28.8 & -30.3 &  17.9 & \nodata & \nodata & \nodata & \nodata & \nodata & \nodata  \\ 
 1 & 566234 &    BL2+? & 17.6 & \nodata &   0.6 &  49.6 &  35.2 &  44.7 &  -0.6 &  36.0 & \nodata & \nodata & \nodata & \nodata  \\ 
 1 & 587412 &      BL2 & 14.8 & \nodata &  14.6 &  38.2 & -14.8 &  26.1 & \nodata & \nodata & \nodata & \nodata & \nodata & \nodata  \\ 
 1 &  36798 &      BL2 &  4.2 & \nodata &  16.5 &  16.0 & -16.3 &   5.3 & \nodata & \nodata & \nodata & \nodata & \nodata & \nodata  \\ 
 1 & 421285 &      BL2 &  9.9 & \nodata &  17.2 &  20.9 & -17.1 &  19.0 & \nodata & \nodata & \nodata & \nodata & \nodata & \nodata  \\ 
 1 & 608436 &      BL1 &  4.9 &   2.3 &  21.5 &  24.4 & \nodata & \nodata & \nodata & \nodata & \nodata & \nodata & \nodata & \nodata  \\ 
 1 & 431836 &      BL1 & 12.1 &   5.5 &  14.4 &  18.7 & \nodata & \nodata & \nodata & \nodata & \nodata & \nodata & \nodata & \nodata  \\ 
 1 &  59700 &       PC & 22.2 & \nodata &  -0.7 &  29.7 & \nodata & \nodata & \nodata & \nodata & \nodata & \nodata & \nodata & \nodata  \\ 
 1 &  59431 &      BL2 & 14.8 & \nodata &   9.1 &  33.5 &  -9.2 &  24.6 & \nodata & \nodata & \nodata & \nodata & \nodata & \nodata  \\ 
 1 & 619252 &       PC & 24.5 & \nodata &  -0.6 &  13.9 & \nodata & \nodata & \nodata & \nodata & \nodata & \nodata & \nodata & \nodata  \\ 
 1 & 269030 &       PC & 50.7 & \nodata &   0.6 & 112.2 & \nodata & \nodata & \nodata & \nodata & \nodata & \nodata & \nodata & \nodata  \\ 
 1 & 452510 &      BL1 &  5.1 &   5.1 &  45.9 &  12.8 & \nodata & \nodata & \nodata & \nodata & \nodata & \nodata & \nodata & \nodata  \\ 
 1 & 652490 &       PC &  5.1 & \nodata &   0.5 &   4.8 & \nodata & \nodata & \nodata & \nodata & \nodata & \nodata & \nodata & \nodata  \\ 
 1 & 474747 &      BL2 & 11.8 & \nodata &  15.8 &  20.7 & -16.0 &  16.5 & \nodata & \nodata & \nodata & \nodata & \nodata & \nodata  \\ 
 1 & 665317 &      BL1 & 14.7 &  14.7 & -22.5 &  29.3 & \nodata & \nodata & \nodata & \nodata & \nodata & \nodata & \nodata & \nodata  \\ 
 1 & 130718 &      BL? &  6.5 & \nodata &  14.5 &  23.7 &  18.0 &  16.3 & \nodata & \nodata & \nodata & \nodata & \nodata & \nodata  \\ 
 1 & 678325 &       PC &  5.8 & \nodata &   0.5 &   6.7 & \nodata & \nodata & \nodata & \nodata & \nodata & \nodata & \nodata & \nodata  \\ 
 1 & 366334 &      BL2 & 26.1 & \nodata & -12.9 &  46.0 &  13.0 &  31.1 & \nodata & \nodata & \nodata & \nodata & \nodata & \nodata  \\ 
 1 & 366819 &      BL2 & 18.5 & \nodata &  22.4 &  40.7 & -22.3 &  33.1 & \nodata & \nodata & \nodata & \nodata & \nodata & \nodata  \\ 
 2 & 436573 &      BL2 & 20.0 & \nodata &  -6.1 &  45.1 &   6.3 &  29.3 & \nodata & \nodata & \nodata & \nodata & \nodata & \nodata  \\ 
 2 & 447803 &      BL? &  9.7 & \nodata &  -3.6 &  13.5 &  15.3 &  11.4 & \nodata & \nodata & \nodata & \nodata & \nodata & \nodata  \\ 
 2 &  40061 &       PC & 26.9 & \nodata &  -0.7 &  28.6 & \nodata & \nodata & \nodata & \nodata & \nodata & \nodata & \nodata & \nodata  \\ 
 2 & 459476 &      BL1 &  5.7 &   3.3 &  32.9 &  16.9 & \nodata & \nodata & \nodata & \nodata & \nodata & \nodata & \nodata & \nodata  \\ 
 2 & 279494 &      BL2 &  5.1 & \nodata & -17.8 &  10.9 &  17.8 &   6.4 & \nodata & \nodata & \nodata & \nodata & \nodata & \nodata  \\ 
 2 & 304657 &      BL2 & 11.3 & \nodata &  -3.4 &  19.6 &   3.2 &  18.4 & \nodata & \nodata & \nodata & \nodata & \nodata & \nodata  \\ 
 2 & 518211 &      BL1 &  3.4 &   3.6 &  16.5 &  19.2 & \nodata & \nodata & \nodata & \nodata & \nodata & \nodata & \nodata & \nodata  \\ 
 2 & 318604 &      BL2 & 19.1 & \nodata &  23.5 &  52.7 & -23.3 &  37.6 & \nodata & \nodata & \nodata & \nodata & \nodata & \nodata  \\ 
 2 & 530487 &      BL1 & 18.0 &  18.0 & -34.0 &  30.9 & \nodata & \nodata & \nodata & \nodata & \nodata & \nodata & \nodata & \nodata  \\ 
 2 & 722091 &      BL1 &  7.8 &   6.1 &  20.7 &  22.9 & \nodata & \nodata & \nodata & \nodata & \nodata & \nodata & \nodata & \nodata  \\ 
 2 & 733835 &      BL1 & 54.7 &  54.7 &  -2.3 &  38.0 & \nodata & \nodata & \nodata & \nodata & \nodata & \nodata & \nodata & \nodata  \\ 
 2 & 135403 &      BL? & 15.4 & \nodata &  49.6 &  32.4 &   1.1 &  19.8 & \nodata & \nodata & \nodata & \nodata & \nodata & \nodata  \\ 
 2 & 345581 &       PC &  5.5 & \nodata &   0.5 &   6.8 & \nodata & \nodata & \nodata & \nodata & \nodata & \nodata & \nodata & \nodata  \\ 
 2 & 542801 &      BL2 &  7.2 & \nodata &  38.7 &  14.4 & -38.8 &  11.1 & \nodata & \nodata & \nodata & \nodata & \nodata & \nodata  \\ 
 2 & 733728 &      BL? & 53.6 & \nodata &  25.1 &  73.9 &  -0.8 &  52.2 & \nodata & \nodata & \nodata & \nodata & \nodata & \nodata  \\ 
 2 & 555864 &      BL1 &  6.1 &   3.3 &   6.3 &  19.6 & \nodata & \nodata & \nodata & \nodata & \nodata & \nodata & \nodata & \nodata  \\ 
 2 & 581017 &   BL2+PC & 15.0 & \nodata &  17.7 &  30.9 &   0.6 &  12.0 & -17.4 &  10.4 & \nodata & \nodata & \nodata & \nodata  \\ 
 2 & 783422 &      BL2 & 15.6 & \nodata &  12.9 &  25.3 & -13.1 &  21.4 & \nodata & \nodata & \nodata & \nodata & \nodata & \nodata  \\ 
 3 & 422608 &      BL1 & 17.1 &  17.1 &  25.1 &  46.4 & \nodata & \nodata & \nodata & \nodata & \nodata & \nodata & \nodata & \nodata  \\ 
 3 & 216589 &      BL2 & 20.7 & \nodata &   9.6 &  36.2 & -11.9 &  20.3 & \nodata & \nodata & \nodata & \nodata & \nodata & \nodata  \\ 
 3 & 422817 &      BL1 &  5.9 &   4.7 &  14.2 &   9.8 & \nodata & \nodata & \nodata & \nodata & \nodata & \nodata & \nodata & \nodata  \\ 
\enddata

%% file: tab4.tex
\startdata
 1 & 0.9791 & -3.6962 &  37 &  46 & 0.516$\pm$0.012 & 0.519$\pm$0.011 &     0.068 &    0.837 \\ 
 2 & 2.1339 & -3.5333 &  30 &  39 & 0.505$\pm$0.010 & 0.511$\pm$0.009 &     0.046 &    0.760 \\ 
 3 & 0.0064 & -2.0142 &  65 & 117 & 0.535$\pm$0.011 & 0.530$\pm$0.011 &     0.090 &    1.385 \\ 
 4 & 0.3235 & -2.0853 &  87 & 122 & 0.510$\pm$0.008 & 0.511$\pm$0.008 &     0.074 &    1.271 \\ 
 5 & -0.3321 & -1.4061 &   0 &  72 &         \nodata &         \nodata &     0.000 &  \nodata \\ 
 6 & -0.3509 & -5.7804 &  14 &  20 & 0.562$\pm$0.012 & 0.549$\pm$0.012 &     0.043 &    0.697 \\ 
 7 & -0.2358 & -5.9869 &  13 &  17 & 0.561$\pm$0.015 & 0.548$\pm$0.014 &     0.047 &    0.691 \\ 
 8 & 10.3770 & -3.8657 &   9 &  15 & 0.629$\pm$0.023 & 0.622$\pm$0.024 &     0.064 &    1.076 \\ 
 9 & 10.4780 & -4.0589 &   7 &  13 & 0.552$\pm$0.035 & 0.549$\pm$0.032 &     0.072 &    1.015 \\ 
10 & 9.5332 & -3.5262 &   5 &  10 & 0.514$\pm$0.035 & 0.517$\pm$0.039 &     0.068 &    1.100 \\ 
11 & 9.6410 & -3.7305 &  10 &  15 & 0.612$\pm$0.045 & 0.605$\pm$0.043 &     0.121 &    1.143 \\ 
12 & 7.7018 & -3.4585 &   7 &  15 & 0.569$\pm$0.046 & 0.566$\pm$0.046 &     0.102 &    1.180 \\ 
13 & 7.8057 & -3.6462 &  14 &  21 & 0.525$\pm$0.025 & 0.523$\pm$0.022 &     0.077 &    1.074 \\ 
14 & 5.1346 & 2.7437 &  14 &  29 & 0.505$\pm$0.014 & 0.489$\pm$0.011 &     0.038 &    1.320 \\ 
15 & 5.2809 & 2.5446 &  16 &  32 & 0.546$\pm$0.017 & 0.540$\pm$0.016 &     0.061 &    1.398 \\ 
16 & 4.9966 & -3.3797 &  13 &  25 & 0.548$\pm$0.017 & 0.542$\pm$0.021 &     0.068 &    1.101 \\ 
17 & 5.1793 & -3.5223 &  17 &  25 & 0.517$\pm$0.018 & 0.512$\pm$0.017 &     0.067 &    0.987 \\ 
18 & 3.8671 & -3.2227 &  20 &  32 & 0.506$\pm$0.016 & 0.510$\pm$0.014 &     0.061 &    0.883 \\ 
19 & 3.9736 & -3.4133 &   5 &  10 & 0.607$\pm$0.069 & 0.595$\pm$0.068 &     0.118 &    0.855 \\ 
20 & 1.5724 & -2.5531 &  44 &  65 & 0.534$\pm$0.013 & 0.531$\pm$0.012 &     0.080 &    0.984 \\ 
21 & 1.7038 & -2.7437 &  27 &  45 & 0.524$\pm$0.013 & 0.519$\pm$0.013 &     0.066 &    0.926 \\ 
22 & -0.3622 & -3.0323 &  27 &  61 & 0.515$\pm$0.018 & 0.520$\pm$0.016 &     0.082 &    1.164 \\ 
23 & -0.6022 & -3.4458 &  27 &  46 & 0.516$\pm$0.012 & 0.515$\pm$0.012 &     0.058 &    1.285 \\ 
24 & -2.5396 & -3.4374 &  28 &  38 & 0.525$\pm$0.012 & 0.522$\pm$0.011 &     0.058 &    1.284 \\ 
25 & -2.4201 & -3.6441 &  27 &  39 & 0.525$\pm$0.012 & 0.523$\pm$0.012 &     0.058 &    1.183 \\ 
26 & -5.0003 & -3.4381 &  27 &  37 & 0.522$\pm$0.015 & 0.525$\pm$0.014 &     0.072 &    0.980 \\ 
27 & -5.0240 & -3.7231 &  12 &  18 & 0.523$\pm$0.028 & 0.521$\pm$0.028 &     0.088 &    0.851 \\ 
28 & -6.8543 & -4.5107 &   0 &   0 &         \nodata &         \nodata &     0.000 &  \nodata \\ 
29 & -6.7370 & -4.6923 &  11 &  16 & 0.554$\pm$0.016 & 0.542$\pm$0.016 &     0.047 &    0.787 \\ 
30 & 1.8436 & -2.9276 &  49 &  67 & 0.525$\pm$0.014 & 0.519$\pm$0.014 &     0.095 &    0.973 \\ 
31 & 2.1299 & -3.0077 &  27 &  51 & 0.523$\pm$0.015 & 0.523$\pm$0.014 &     0.069 &    0.894 \\ 
32 & 2.2353 & -3.2179 &  30 &  40 & 0.519$\pm$0.016 & 0.521$\pm$0.016 &     0.086 &    0.827 \\ 
33 & 2.2411 & -3.7339 &  35 &  50 & 0.548$\pm$0.015 & 0.544$\pm$0.015 &     0.087 &    0.855 \\ 
34 & 1.2466 & -2.4852 &  51 &  68 & 0.517$\pm$0.014 & 0.517$\pm$0.013 &     0.092 &    1.181 \\ 
35 & 2.9448 & -3.0807 &  23 &  36 & 0.532$\pm$0.016 & 0.530$\pm$0.014 &     0.064 &    0.927 \\ 
36 & 3.0602 & -3.2802 &  31 &  41 & 0.523$\pm$0.012 & 0.520$\pm$0.012 &     0.062 &    0.823 \\ 
37 & -0.1047 & -1.8190 &   7 &  98 & 0.560$\pm$0.037 & 0.564$\pm$0.041 &     0.092 &    1.516 \\ 
38 & 0.8742 & -3.4880 &  40 &  47 & 0.537$\pm$0.007 & 0.532$\pm$0.008 &     0.050 &    0.917 \\ 
39 & 0.4221 & -2.2934 &  56 &  89 & 0.527$\pm$0.012 & 0.525$\pm$0.011 &     0.083 &    1.297 \\ 
40 & -3.0962 & -3.2304 &  20 &  38 & 0.500$\pm$0.020 & 0.506$\pm$0.019 &     0.080 &    1.417 \\ 
41 & -2.8799 & -3.3503 &  24 &  45 & 0.538$\pm$0.015 & 0.535$\pm$0.017 &     0.078 &    1.309 \\ 
42 & 4.3910 & -3.4491 &  19 &  28 & 0.552$\pm$0.028 & 0.557$\pm$0.026 &     0.108 &    1.152 \\ 
43 & 0.2705 & 2.8736 &   3 &  51 & 0.564$\pm$0.057 & 0.550$\pm$0.046 &     0.046 &    1.601 \\ 
44 & -0.5268 & -1.2732 &   0 &  28 &         \nodata &         \nodata &     0.000 &  \nodata \\ 
45 & 0.9857 & -3.9573 &  38 &  43 & 0.498$\pm$0.010 & 0.494$\pm$0.009 &     0.053 &    0.828 \\ 
46 & 1.0943 & -4.1578 &  32 &  42 & 0.530$\pm$0.011 & 0.524$\pm$0.010 &     0.057 &    0.898 \\ 
47 & -11.2050 & -2.6179 &   5 &   5 & 0.517$\pm$0.022 & 0.507$\pm$0.030 &     0.052 &    1.235 \\ 
48 & -11.0759 & -2.8024 &   1 &   3 & 0.647$\pm$0.080 & 0.627$\pm$0.080 &   \nodata &    1.018 \\ 
49 & -11.3595 & -3.2618 &   2 &   3 & 0.593$\pm$0.033 & 0.596$\pm$0.010 &     0.010 &    0.881 \\ 
\enddata